\documentclass[prl,10pt,twocolumn,superscriptaddress,aps,floatfix]{revtex4-1}
\usepackage{graphicx}
\usepackage{natbib}
\usepackage{url}
\usepackage[normalem]{ulem}
\usepackage{color}
\usepackage{amsmath}

\begin{document}


\title{Quantization of Hall Resistance at the Metallic Interface between an Oxide Insulator and SrTiO$_{3}$}
\author{Felix Trier}%
\email{To whom correspondence should be adressed: \\fetri@dtu.dk}
\affiliation{%
	Department of Energy Conversion and Storage, Technical University of Denmark, Ris{\o} Campus, 4000 Roskilde, Denmark.
}%
\author{Guenevere E.D.K. Prawiroatmodjo}
\affiliation{%
 Center for Quantum Devices, Niels Bohr Institute, University of Copenhagen, Universitetsparken 5, 2100 Copenhagen, Denmark.
}%
\author{Zhicheng Zhong}
\affiliation{%
	Max Planck Institute for Solid State Research, D-70569 Stuttgart, Germany.
}%
\author{Dennis Valbj{\o}rn Christensen}%
\affiliation{%
	Department of Energy Conversion and Storage, Technical University of Denmark, Ris{\o} Campus, 4000 Roskilde, Denmark.
}%
\author{Merlin von Soosten}%
\affiliation{%
	Department of Energy Conversion and Storage, Technical University of Denmark, Ris{\o} Campus, 4000 Roskilde, Denmark.
}%
\affiliation{%
	Center for Quantum Devices, Niels Bohr Institute, University of Copenhagen, Universitetsparken 5, 2100 Copenhagen, Denmark.
}%
\author{Arghya Bhowmik}%
\affiliation{%
	Department of Energy Conversion and Storage, Technical University of Denmark, Ris{\o} Campus, 4000 Roskilde, Denmark.
}%
\author{Juan Maria Garc\'{i}a Lastra}%
\affiliation{%
	Department of Energy Conversion and Storage, Technical University of Denmark, Ris{\o} Campus, 4000 Roskilde, Denmark.
}%
\author{Yunzhong Chen}%
\email{yunc@dtu.dk}
\affiliation{%
Department of Energy Conversion and Storage, Technical University of Denmark, Ris{\o} Campus, 4000 Roskilde, Denmark.
}%
\author{Thomas Sand Jespersen}%
\affiliation{%
	Center for Quantum Devices, Niels Bohr Institute, University of Copenhagen, Universitetsparken 5, 2100 Copenhagen, Denmark.
}%
\author{Nini Pryds}%
\email{nipr@dtu.dk}
\affiliation{%
Department of Energy Conversion and Storage, Technical University of Denmark, Ris{\o} Campus, 4000 Roskilde, Denmark.
}%
\date{\today}

\begin{abstract}
The two-dimensional metal forming at the interface between an oxide insulator and SrTiO$_{3}$ provides new opportunities for oxide electronics. However, the quantum Hall effect, one of the most fascinating effects of electrons confined in two dimensions, remains underexplored at these complex oxide heterointerfaces. Here, we report the experimental observation of quantized Hall resistance in a SrTiO$_{3}$ heterointerface based on the modulation-doped amorphous-LaAlO$_{3}$/SrTiO$_{3}$ heterostructure, which exhibits both high electron mobility exceeding 10,000 cm$^2$/V\,s and low carrier density on the order of $\sim$10$^{12}$ cm$^{-2}$. Along with unambiguous Shubnikov--de Haas oscillations, the spacing of the quantized Hall resistance suggests that the interface is comprised of a single quantum well with ten parallel conducting two-dimensional sub-bands. This provides new insight into the electronic structure of conducting oxide interfaces and represents an important step towards designing and understanding advanced oxide devices.
\end{abstract}

\maketitle

The quantum Hall effect (QHE), arises from quantization of the cyclotron motion of charge carriers subjected to a perpendicular magnetic field. The cyclotron motion becomes quantized when electrons complete enclosed orbits without being scattered, and the observation of the QHE therefore requires materials with high carrier mobility. Consequently, it has only been observed in a few material systems such as semiconductor heterostructures based on silicon \cite{klitzing_new_1980} or III-V compounds \cite{tsui_two-dimensional_1982}, Bi$_{2}$Se$_{3}$ \cite{cao_quantized_2012} and graphene \cite{zhang_experimental_2005}. Recent technical advances in the growth of oxides have resulted in the creation of high-mobility two-dimensional electron gases (2-DEGs) at heterointerfaces based on either ZnO \cite{tsukazaki_quantum_2007, tsukazaki_observation_2010} or SrTiO$_{3}$ (STO) \cite{ohtomo_high-mobility_2004, chen_high-mobility_2013}. Remarkably, the mobility enhancement made in polar MgZnO/ZnO heterostructures has led to the observation of both integer and fractional QHE \cite{tsukazaki_quantum_2007, tsukazaki_observation_2010, falson_even-denominator_2015}. The conducting states of ZnO-based heterostructures are, however, similar to the conventional semiconductor heterostructures, derived from \textit{sp} hybrid orbitals with a covalent bond nature. In contrast, the interface conductivity in STO-based heterostructures originates from less overlapping Ti 3\textit{d} orbitals, where the resulting ionic bonds lead to a strong coupling between the lattice, charge and spin degrees of freedom. The strong electronic correlations give rise to a variety of properties, such as gate-tunable superconductivity, magnetism, and tunable metal-insulator transitions \cite{thiel_tunable_2006, christensen_controlling_2013, reyren_superconducting_2007, caviglia_electric_2008, brinkman_magnetic_2007}, which has generated particular interest in STO-based heterostructures \cite{sulpizio_nanoscale_2014}.

\begin{figure}
 	\includegraphics[width=8.5cm]{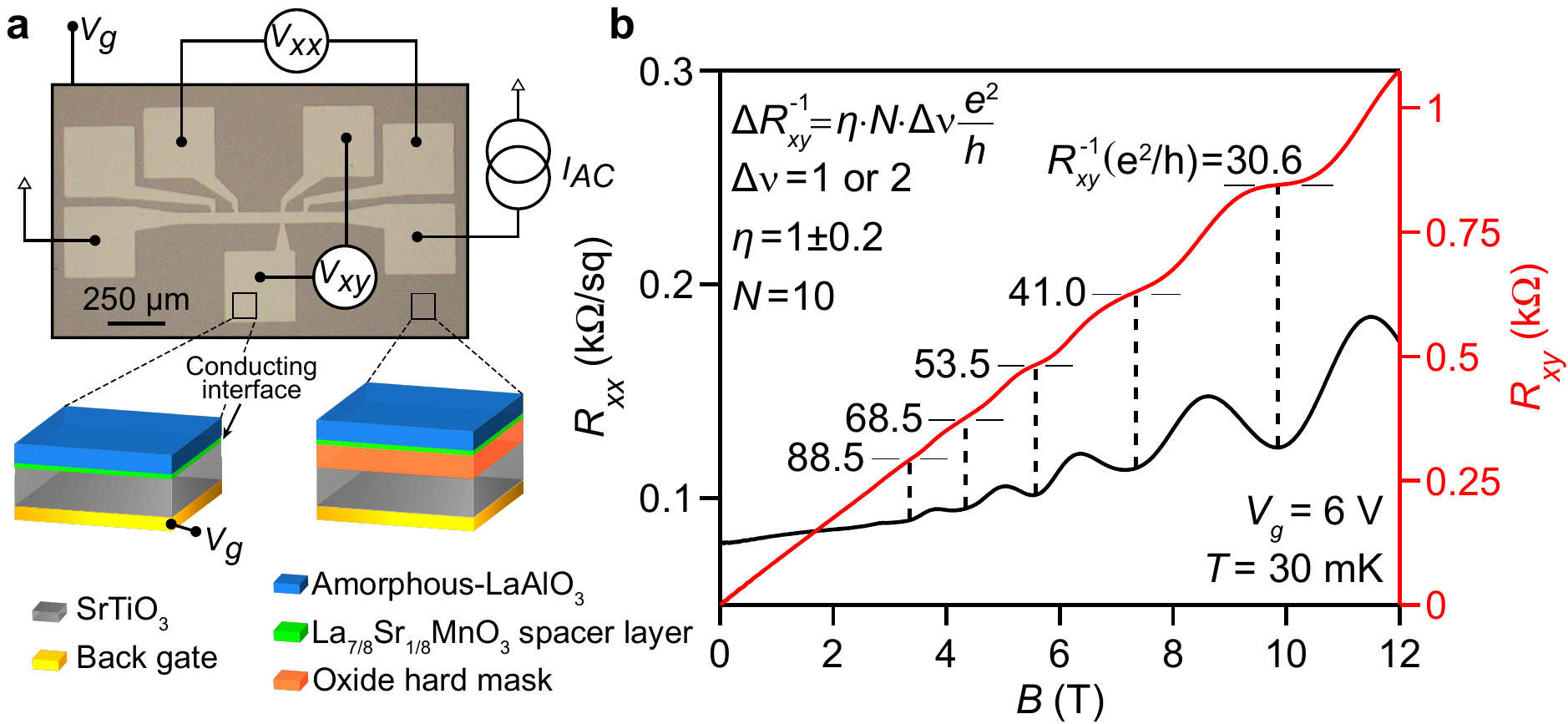}
 	\caption{a, Optical microscopy image of the Hall bar device where the electrostatic doping is controlled by a sample-wide back gate (\textit{V$_{g}$}). b, \textit{R$_{xx}$} and \textit{R$_{xy}$} as a function of magnetic field (\textit{B}) normal to the sample surface. The vertical dashed lines correspond to minima in \textit{dR$_{xy}$}/\textit{dB}.}
 	\label{figure1}
\end{figure}
 
The metallic conductivity of STO-based heterostructures originates from degenerate Ti 3\textit{d} \textit{t$_{2g}$} levels \cite{delugas_spontaneous_2011, son_density_2009, khalsa_theory_2012, breitschaft_two-dimensional_2010} with the orbital occupation strongly depending on the carrier density. Below a threshold carrier density of 1.7$\times$10$^{13}$ cm$^{-2}$ in the intensively investigated LaAlO$_{3}$/SrTiO$_{3}$ (LAO/STO) interface \cite{joshua_universal_2012}, electrons are confined within a thin sheet of STO in Ti 3\textit{d$_{xy}$} levels that are characterized by strong coupling parallel to the interface. Above this threshold density, electrons begin to populate the 3\textit{d$_{xz/yz}$} levels, which extend further away from the interface and add out-of-plane coupling to the electronic system \cite{gabay_oxide_2013}. The Fermi surfaces of STO-based heterointerfaces have been probed by Shubnikov--de Haas (SdH) oscillations in the longitudinal resistance, \textit{R$_{xx}$} \cite{chen_high-mobility_2013, caviglia_two-dimensional_2010, ben_shalom_shubnikovhaas_2010, xie_quantum_2014, mccollam_quantum_2014, son_epitaxial_2010}. Notably, the carrier density deduced from SdH oscillations, \textit{n$_{s}^{\mathrm{SdH}}$}, is often more than 1 order of magnitude lower than that obtained by the Hall effect, \textit{n$_{s}^{\mathrm{Hall}}$}. This discrepancy has been ascribed to the hypothesis that only a small fraction of the carriers contributes to the SdH oscillations \cite{chen_high-mobility_2013, caviglia_two-dimensional_2010, ben_shalom_shubnikovhaas_2010, xie_quantum_2014, mccollam_quantum_2014, son_epitaxial_2010}. Although the two-dimensional nature of the interface conductivity is demonstrated by the dependence of the SdH oscillations on magnetic field tilt angle, only a few studies report plateaulike structures in the Hall resistance, \textit{R$_{xy}$} \cite{xie_quantum_2014, mccollam_quantum_2014}. Full analysis of the quantizationlike Hall resistance in these studies was, however, difficult due to irregular SdH oscillation spectra \cite{xie_quantum_2014, mccollam_quantum_2014}. Consequently, a clear QHE with well-defined quantized Hall plateaus coinciding with SdH oscillation minima remains undemonstrated at complex oxide interfaces. Here, we report unambiguous experimental observation of quantized Hall resistance in complex oxides (see Fig. 1b). Remarkably, the inverse Hall resistance (\textit{R$_{xy}^{-1}$}) is found to be regularly spaced for all plateaus following the relation $\Delta$\textit{R$_{xy}^{-1}$} = $\eta N\Delta\nu\frac{e^2}{h}$, with \textit{h} being Planck's constant, \textit{e} the electron charge, $\Delta \nu$ the filling factor step, \textit{N} a scaling factor equaling 10, and $\eta = 1.0\pm0.2$. This manifestation of the QHE is different from what is usually observed in conventional semiconductors, MgZnO/ZnO or other two-dimensional materials \cite{klitzing_new_1980, tsui_two-dimensional_1982, cao_quantized_2012, zhang_experimental_2005, tsukazaki_quantum_2007, tsukazaki_observation_2010, falson_even-denominator_2015}, highlighting the exotic nature and potential opportunities of complex oxide devices. As will become clear later, our results are consistent with the oxide interface being comprised of a single quantum well hosting ten parallel conducting Ti 3\textit{d$_{xy}$} sub-bands, which is in contrast to the usual situation in LAO/STO for \textit{n$_{s}^{\mathrm{Hall}}$} $>$ 1.7$\times$10$^{13}$ cm$^{-2}$ where all \textit{t$_{2g}$} levels (Ti 3\textit{d$_{xy}$} and 3\textit{d$_{xz/yz}$} sub-bands) are occupied.

The modulation-doped oxide interface is realized by inserting a La$_{7/8}$Sr$_{1/8}$MnO$_{3}$ (LSM) spacer layer at the interface between amorphous LAO (a-LAO) and the crystalline STO substrate (a-LAO/LSM/STO) \cite{chen_extreme_2015}. A Hall bar device is fabricated (see Fig. 1a) by conventional electron-beam lithography and an oxide hard mask of a-LSM \cite{trier_patterning_2015} rather than a-LAO \cite{schneider_microlithography_2006}. Standard ac lock-in measurements (\textit{I$_{ac}$} = 10 nA, \textit{f$_{ac}$} = 77.03 Hz) are performed in a dilution refrigerator at temperatures between 30 mK and 3 K, back gate potentials (\textit{V$_g$}) between $-5$ and 10 V, and magnetic fields between 0 and 12 T. Sample rotation between 0$^{\circ}$ and $\sim$82$^{\circ}$ is achieved by a piezoelectric rotator. SdH oscillations are fitted by the analytical expression from Coleridge \textit{et al.} \cite{coleridge_low-field_1989}. Density functional theory-based tight binding (DFT-TB) calculations follow the approach outlined by Zhong \textit{et al.} \cite{zhong_quantum_2013}.

The Hall bar device shows metallic conductivity similar to the unpatterned interface \cite{chen_extreme_2015} with a carrier density \textit{n$_{s}^{\mathrm{Hall}}$} = 1/($e$\textit{R$_H$}) $\sim$5.6$\times$10$^{12}$ cm$^{-2}$, where \textit{R$_H$} = \textit{dR$_{xy}$}/\textit{dB} is the Hall coefficient, and a Hall carrier mobility, $\mu_{\mathrm{Hall}}$ = 1/(e${\cdot}$\textit{n$_{s}^{\mathrm{Hall}}$}${\cdot}$\textit{R$_{xx}$}) $\sim$8,703 cm$^{2}$/V\,s, at 2 K and \textit{V$_g$} = 0 V. This mobility is among the highest reported values for patterned complex oxide interfaces, which is typically below $\sim$1000 cm$^{2}$/V\,s at 2 K \cite{schneider_microlithography_2006, stornaiuolo_-plane_2012}. Importantly, such an enhanced mobility, along with a low carrier density, enables us to observe the quantization of Hall resistance. Note that \textit{R$_{xx}$}(\textit{B}) $\sim$ \textit{R$_{xx}$}(\textit{$-$B}) and \textit{R$_{xy}$}(\textit{B}) $\sim$ $-$\textit{R$_{xy}$}(\textit{$-$B}) [Supplemental Material (SM) Fig. S1]: therefore, the subsequent measurements only concern positive magnetic field direction. In order to examine the Fermi surface of the system based on our initial observation of the SdH oscillations \cite{chen_extreme_2015}, we further investigated the dependence of \textit{R$_{xx}$}(\textit{B}) and \textit{R$_{xy}$}(\textit{B}) on temperature, magnetic field tilt angle, and \textit{V$_g$}.

\begin{figure}
	\includegraphics[width=8.5cm]{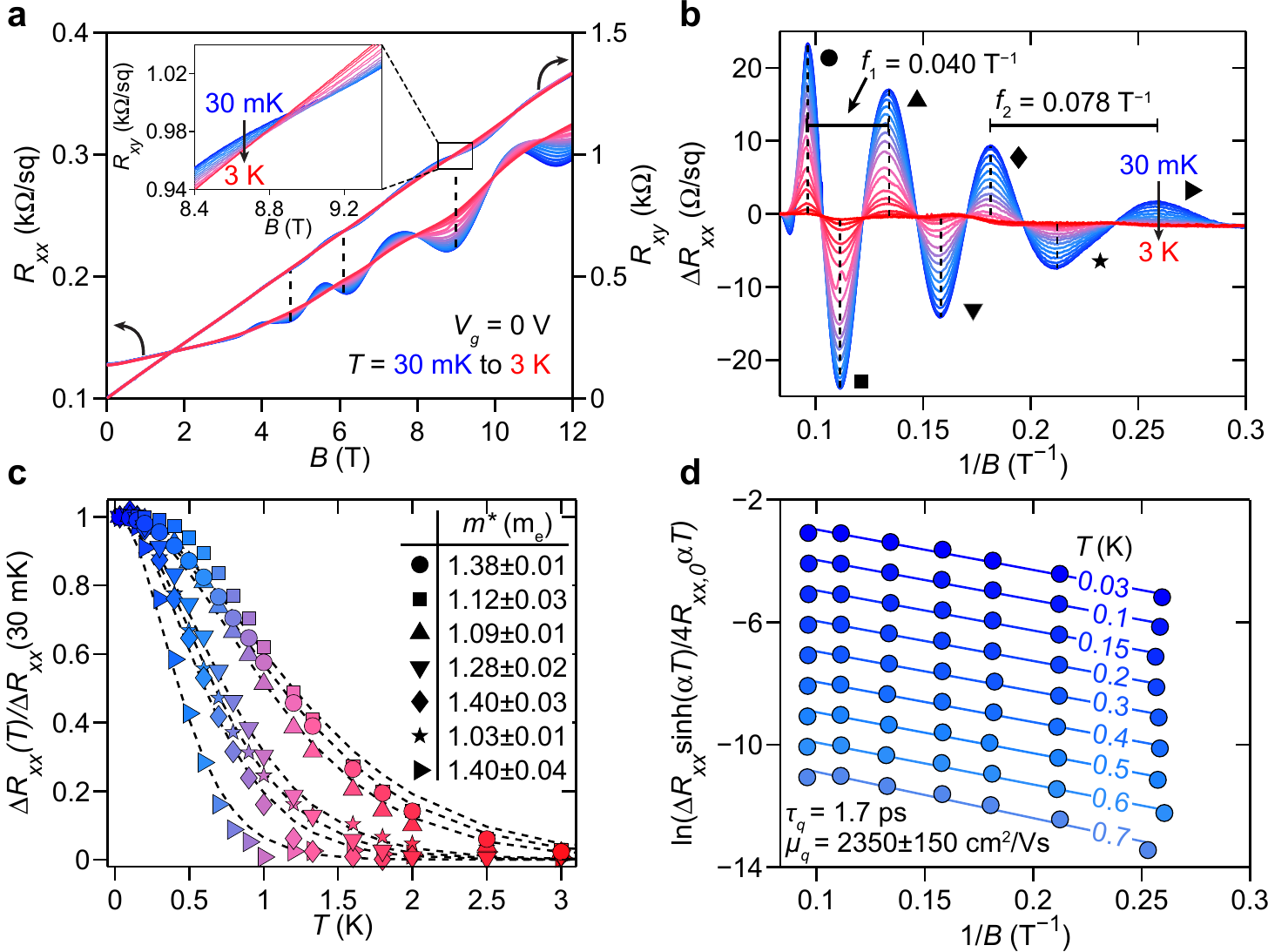}
	\caption{a, \textit{R$_{xx}$} and \textit{R$_{xy}$} as a function of \textit{B} normal to the sample surface. The vertical dashed lines correspond to minima in \textit{dR$_{xy}$}/\textit{dB} at 30 mK. b, Temperature dependence of $\Delta$\textit{R$_{xx}$} as a function of 1/\textit{B}. Vertical dashed lines correspond to the position of oscillation extrema in $\Delta$\textit{R$_{xx}$} at 30 mK. c, Extraction of carrier effective mass (\textit{m$^*$}). d, Dingle plots for \textit{T} = 30 mK to 0.7 K.}
	\label{figure2}
\end{figure}

Figure 2a shows the temperature dependence of \textit{R$_{xx}$}(\textit{B}) and \textit{R$_{xy}$}(\textit{B}) for \textit{V$_g$} = 0 V and Fig. 2b shows \textit{R$_{xx}$} subtracted a background (SM Fig. S2), $\Delta$\textit{R$_{xx}$}, as a function of 1/\textit{B} for different temperatures. The amplitudes of the SdH oscillations clearly reduce upon heating and the carrier effective mass, \textit{m$^*$}, is extracted by fitting the temperature dependent term in the Ando relation \cite{ando_electronic_1982}: $\Delta$\textit{R$_{xx}$}(\textit{T})/$\Delta$\textit{R$_{xx}$}(\textit{T$_0$}) = \textit{T}sinh($\alpha$\textit{T$_0$})/\textit{T$_0$}sinh($\alpha$\textit{T}), where \textit{T$_0$} = 30 mK, $\alpha$ = 2$\pi^2k_B$/$\hbar\omega_c$, $\omega_c$ = \textit{eB}/\textit{m$^*$}, and $k_B$ is the Boltzmann constant (Fig. 2c). The effective mass, \textit{m$^*$}, is found to be between 1.03 and 1.40 m$_\textrm{e}$, similar to typical values reported for all-crystalline LAO/STO interfaces \cite{caviglia_two-dimensional_2010, ben_shalom_shubnikovhaas_2010, xie_quantum_2014}. Also apparent in Fig. 2b are two different 1/\textit{B} frequencies, \textit{f$_1$} = 0.040 T$^{-1}$ and \textit{f$_2$} = 0.078 T$^{-1}$ differing by a factor of $\sim$2 consistent with a transition from an apparent spin degenerate to a fully spin resolved electronic system at \textit{B} $\approx$ 6 T (SM Fig. S3). Note that this apparent spin degeneracy is different from that found in conventional semiconductors due to the much larger effective mass in oxides leading to a Landau level (LL) splitting, $\hbar\omega_c$, comparable with the Zeeman splitting, $g\mu_B B$ ($g$ is the \textit{g} factor and $\mu_B$ the Bohr magneton). Using a \textit{g} factor \cite{laguta_photoinduced_2002, caviglia_tunable_2010} of 2, the Zeeman and LL splitting energies (for \textit{m$^*$} = 1.30 m$_\textrm{e}$) at \textit{B} = 6 T become 0.70 and 0.54 meV, respectively. In this case, the intrinsic spin degeneracy should be lifted at all fields where SdH oscillations are observed. However, at low magnetic fields, an apparent spin degeneracy occurs when the spin-down part of LL \textit{n} overlaps with the spin-up part of LL \textit{n} + 1 (SM Fig. S3). Thus, the system becomes spin resolved only at fields above $\sim$6 T. For a spin resolved electronic system without valley degeneracy, this leads to a \textit{n$_{s}^{SdH}$} = $e$/(\textit{f$_1$}\textit{h}) $\sim$5.9$\times$10$^{11}$ cm$^{-2}$, which remarkably is $\sim$10 times lower than \textit{n$_{s}^{\mathrm{Hall}}$} similar to other STO-based interfaces \cite{chen_high-mobility_2013, caviglia_two-dimensional_2010, ben_shalom_shubnikovhaas_2010, xie_quantum_2014, mccollam_quantum_2014, son_epitaxial_2010}. The quantum mobility, $\mu_q$, is extracted in Fig. 2d by plotting the 1/\textit{B} dependency of the absolute reduction of the oscillation amplitude using the Dingle formula \cite{coleridge_low-field_1989}: ln[($\Delta$\textit{R$_{xx}$}(\textit{B})sinh($\alpha$\textit{T}))/(4\textit{R$_{xx}$}(\textit{B}=0T)$\alpha$\textit{T})], where $\alpha$ = 2$\pi^2 k_B$/$\hbar\omega_c$, $\omega_c$ = \textit{eB}/\textit{m$^*$}, and \textit{R$_{xx}$}(\textit{B}=0T) = 128 $\Omega$/sq. These Dingle plots reveal a $\mu_q$ of $\sim$2350 cm$^{2}$/V\,s (see Fig. 2d). This quantum mobility is approximately 1 order of magnitude higher than that for all-crystalline LAO/STO interfaces \cite{caviglia_two-dimensional_2010, ben_shalom_shubnikovhaas_2010, xie_quantum_2014}, and comparable to that in the spinel/perovskite $\gamma$-Al$_2$O$_3$/SrTiO$_{3}$ interface \cite{chen_high-mobility_2013}.

\begin{figure}
	\includegraphics[width=8.5cm]{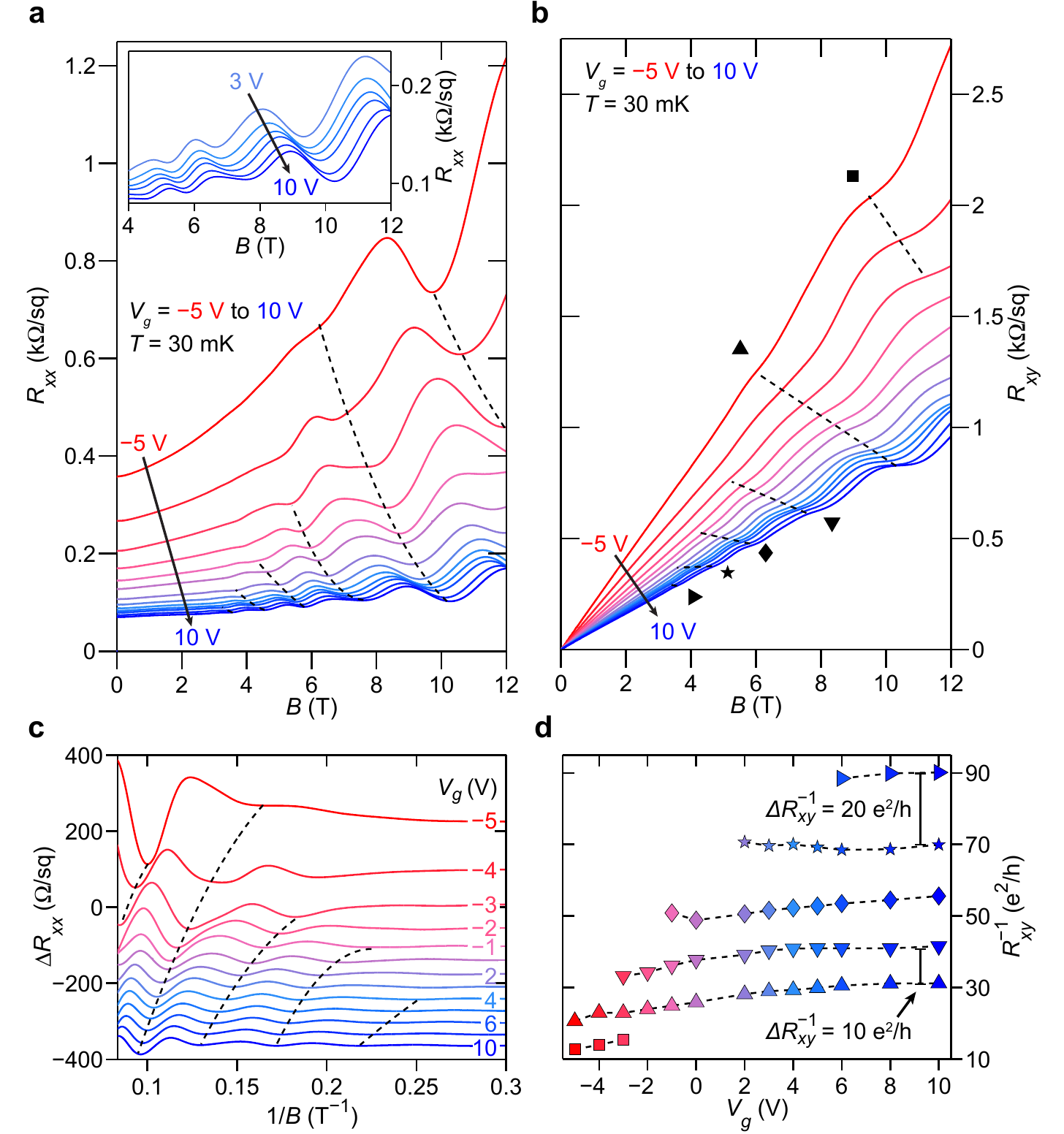}
	\caption{a and b, \textit{R$_{xx}$} and \textit{R$_{xy}$} as a function of \textit{B} normal to the sample surface. Dashed lines correspond to minima in \textit{R$_{xx}$} and \textit{dR$_{xy}$}/\textit{dB}. c, $\Delta$\textit{R$_{xx}$} as a function of 1/\textit{B}. d, \textit{R$_{xy}^{-1}$} in units of $\frac{e^2}{h}$ at different Hall plateaus.}
	\label{figure3}
\end{figure}

When tilting the magnetic field angle the oscillation amplitudes decrease rapidly with the SdH oscillations almost disappearing at angles above $\sim$40$^{\circ}$, consistent with the electronic system being strongly confined at the interface (SM Figs. S4 and S5).

The \textit{R$_{xx}$}(\textit{B}) and \textit{R$_{xy}$}(\textit{B}) traces change significantly as \textit{V$_g$} is varied from $-5$ to 10 V (see Figs. 3a and 3b). Specifically, \textit{n$_{s}^{\mathrm{Hall}}$} changes from 3.0$\times$10$^{12}$ to 7.8$\times$10$^{12}$ cm$^{-2}$ and \textit{$\mu_{\mathrm{Hall}}$} from 5760 to 11,416 cm$^{2}$/V\,s for \textit{V$_g$} = $-5$ V to 10 V (SM Figs. S6a and S6b). For the different \textit{V$_g$}, $\Delta$\textit{R$_{xx}$} shows clear oscillations with a 1/\textit{B} periodicity (see Fig. 3c) and extracting the spin resolved SdH oscillation frequency for the different \textit{V$_g$} results in \textit{n$_{s}^{SdH}$} changing from $\sim$3.8$\times$10$^{11}$ cm$^{-2}$ to 8.5$\times$10$^{11}$ cm$^{-2}$ for \textit{V$_g$} = $-5$ to 10 V (SM Fig. S6c). Notably, \textit{n$_{s}^{\mathrm{Hall}}$} remains $\sim$10${\cdot}$\textit{n$_{s}^{SdH}$} in the full range of investigated \textit{V$_g$} (SM Fig. S6d). Remarkably, the Hall resistance develops a steplike behavior with well-defined quantum Hall plateaus upon increasing \textit{V$_g$} (see Fig. 3b). All Hall resistance plateaus, chosen to be at the minimum position of \textit{dR$_{xy}$}/\textit{dB}, coincide with minima in \textit{R$_{xx}$} and Fig. 3d shows the value of 1/\textit{R$_{xy}$} at the plateaus in units of \textit{e$^{2}$/h} as a function of \textit{V$_g$}. Strikingly, \textit{R$_{xy}^{-1}$} does not appear to assume integer values of \textit{e$^{2}$/h} different from QHE in the conventional semiconductor quantum well comprising a single band where \textit{R$_{xy}^{-1}$} = $\nu\frac{e^2}{h}$ (SM, Sec. 1). Moreover, \textit{R$_{xy}^{-1}$} for the same plateau varies as a function of \textit{V$_g$}, contrary to the case in conventional semiconductors where it remains constant \cite{klitzing_new_1980}. The plateaus, however, appear regularly spaced for all \textit{V$_g$} with either $\Delta R_{xy}^{-1}$ $\sim$10$\pm2$ $e^{2}/h$ or $\Delta R_{xy}^{-1}$ $\sim$20$\pm2$ $e^{2}/h$ for \textit{B} $>$ 6 T and \textit{B} $<$ 6 T, respectively. Assuming a spin resolved or apparent spin degenerate situation where $\Delta\nu$ = 1 or 2, respectively, the inverse Hall resistance at plateaus is spaced by $\Delta R_{xy}^{-1}$ = $\eta$\textit{N}$\Delta\nu\frac{e^2}{h}$, where \textit{N} = 10 and $\eta$ = 1.0$\pm$0.2. The overall linearity of the Hall resistance and the relative flatness of the Hall plateaus at high \textit{V$_g$} and magnetic field suggest that the contribution from a low-mobility parallel conducting channel is negligible. And indeed, the scaling factor $\eta$\textit{N} $\sim$ 10$\pm$2 remains the same when considering the contribution from a low-mobility parallel conducting channel \cite{van_der_burgt_magnetotransport_1995, grayson_measuring_2005} not in the SdH regime (SM, Figs. S7--S9). Xie \textit{et al.} observed quantizationlike Hall resistance at the LAO/STO interface \cite{xie_quantum_2014} with a spacing that scaled with \textit{N} $\sim$4. This was explained by a breaking of the fundamental band symmetry through magnetic breakdown orbits; however, this explanation cannot account for a scaling factor of $\eta$\textit{N} $\sim$ 10$\pm$2 observed here. A consistent description of the data, however, appears if we recognize that the modulation-doped oxide interface may consist of multiple parallel conducting channels that collectively give rise to a combined QHE \cite{cao_quantized_2012, stormer_quantization_1986}. For instance, conduction through highly doped Bi$_2$Se$_3$ occurs through multiple quantum wells, and gives rise to quantized Hall resistance plateaus with \textit{R$_{xy}^{-1}$} = $\eta$\textit{N}$\nu\frac{e^2}{h}$ where $\eta$ = 1.0$\pm$0.2 and \textit{N} is the number of quantum wells \cite{cao_quantized_2012}. However, in that case the quantum wells are almost uncoupled with identical charge carrier densities and almost behave like a classic Kronig-Penney model. On the other hand, due to band bending in STO, the charge carrier density in our parallel electronic channels is expected to vary across the different channels \cite{son_density_2009, khalsa_theory_2012, reich_accumulation_2015} and the \textit{R$_{xy}^{-1}$} quantization is not expected to assume integer values of $\frac{e^2}{h}$. This is a consequence of the measured filling factor, $\nu$(\textit{B}), being an average of the filling factors of each channel, $\nu_i$(\textit{B}), which take integer values at different \textit{B} fields owing to their different carrier densities. Furthermore, the different $\nu_i$(\textit{B}) likewise explain why well-defined plateaus are only observed for certain carrier density combinations where $\nu_i$(\textit{B}) are near multiples of each other (SM, Sec. S1). However, the step size, $\Delta R_{xy}^{-1}$, between plateaus is expected to take integer values of \textit{N}$\frac{e^2}{h}$ (SM, Sec. S1) where \textit{N} is the number of parallel conducting channels.

\begin{figure}
	\includegraphics[width=8.5cm]{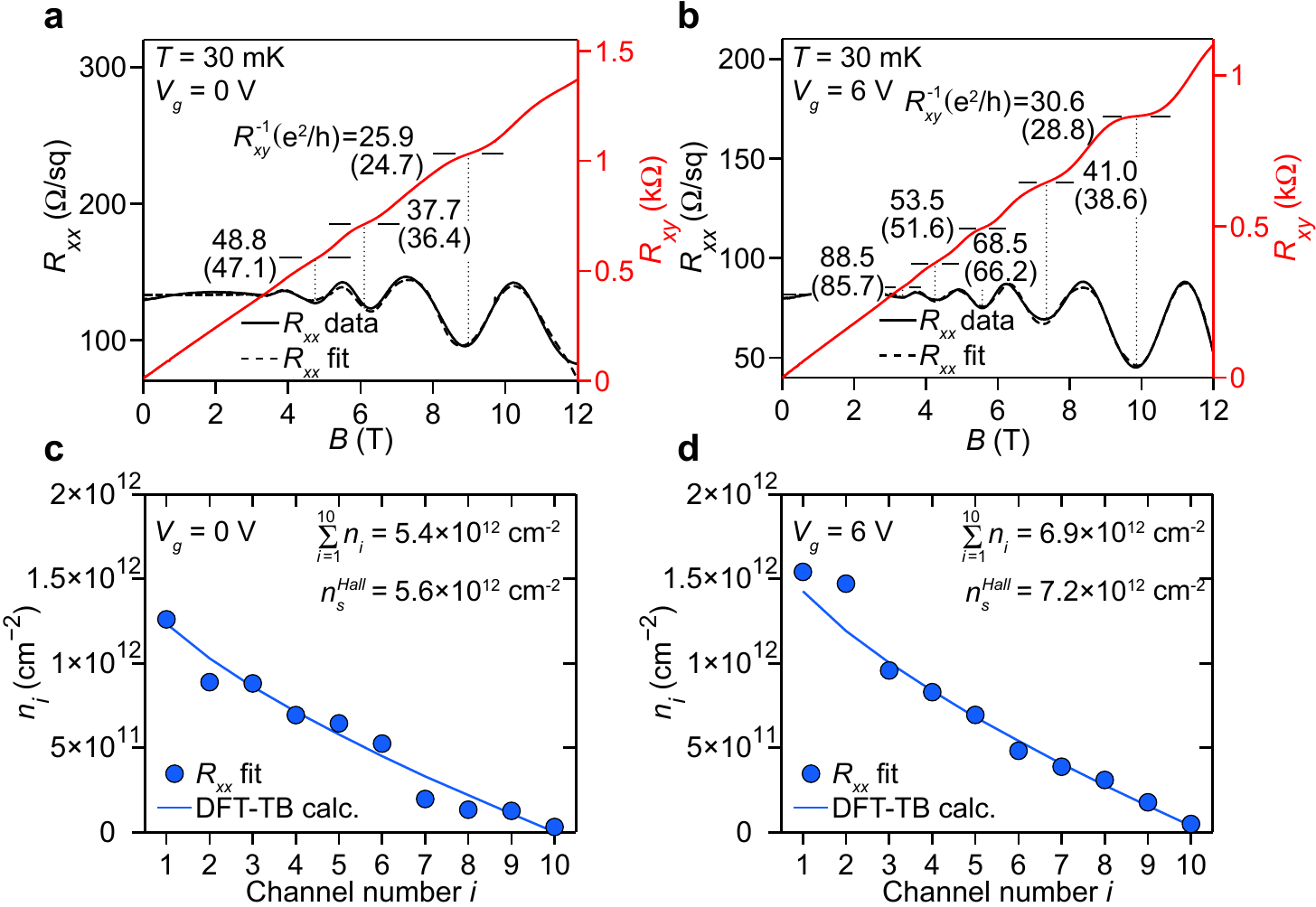}
	\caption{a,b, Measured \textit{R$_{xx}$} data along with the \textit{R$_{xx}$} fit. \textit{R$_{xy}^{-1}$} in units of $\frac{e^2}{h}$ are indicated at plateaus along with calculated values from the fit shown in parenthesis. c,d, The fitted charge carrier densities (\textit{n$_i$}) for the ten parallel channels.}
	\label{figure4}
\end{figure}

In order to quantitatively address whether such parallel electronic channels with varying charge carrier density can account for the observed \textit{R$_{xx}$} oscillations and \textit{R$_{xy}$} plateaus, the data are fitted (SM, Sec. S2 and Figs. S10 and S11) by summing contributions from \textit{N} = 10 parallel conducting channels that are either apparent spin degenerate (for \textit{B}$<$6 T) or spin resolved (for \textit{B}$>$6 T). Here, the analytical expression for \textit{R$_{xx,i}$} oscillations of each channel (\textit{i}=1,2,3,...,10) with charge carrier density \textit{n$_i$} is given by \cite{coleridge_low-field_1989}
\begin{equation*}
\begin{split}
&R_{xx,i}(\textit{B}) = \frac{m_i^*}{n_i e^2 \tau_{0,i}} \\ \times \bigg[&1-4e^{-\pi/\omega_{c,i}\tau_{q,i}} \frac{2\pi^2 k_B T/\hbar \omega_{c,i}}{\textrm{sinh}(2\pi^2 k_B T/\hbar\omega_{c,i})}\textrm{cos}\bigg(2\pi \frac{h n_i}{2e\textit{B}}\bigg) \bigg].
\end{split}
\end{equation*}
For \textit{V$_g$} = 0 and 6 V (see Figs. 4a and 4b), there is an excellent agreement between the measured \textit{R$_{xx}$} oscillations subtracted a magnetoresistance background and the calculated \textit{R$_{xx}$} traces with fitted values of \textit{n$_i$} as illustrated in Figs. 4c and 4d. Furthermore, \textit{R$_{xy}^{-1}$} calculated using the fitted charge carrier densities agrees well with the measured plateau values for both \textit{V$_g$} = 0 and 6 V (Figs. 4a and 4b). Importantly, the obtained set of charge carrier densities consistently predicts not only the observed $\Delta R_{xy}^{-1}$ steps of $\sim$10$\pm$2 $e^{2}/h$ or $\Delta R_{xy}^{-1}$ $\sim$20$\pm$2 $e^{2}/h$ but also the unconventional \textit{R$_{xy}^{-1}$} variation observed in Fig. 3d for the two \textit{V$_g$} considered. Ultimately, we believe that this physical picture is representable for the entire gate-voltage range investigated, i.e., that the ten parallel conducting electronic channels are responsible for the observed \textit{R$_{xx}$} oscillations and \textit{R$_{xy}$} plateaus from \textit{V$_g$} = $-5$ to 10 V. In other words, the modulation-doped oxide interface consists of multiple parallel conducting two-dimensional channels with similar effective mass, Hall mobility, and quantum mobility, whereas the charge carrier density of different channels varies greatly. Upon application of an electric field all channels populate (deplete) concurrently with the same normalized gate dependence \Big($\frac{dn_i}{dV_g}\frac{1}{n_i}$\Big). Moreover, the number of channels is expected to increase (decrease) when the electronic system is populated/depleted sufficiently beyond the investigated gate-voltage range of \textit{V$_g$} = $-5$ to 10 V and will be the subject of future studies. This physical picture is consistent with recent experiments on strongly carrier depleted LAO/STO \cite{pallecchi_giant_2015}.

\begin{figure}
	\includegraphics[width=8.5cm]{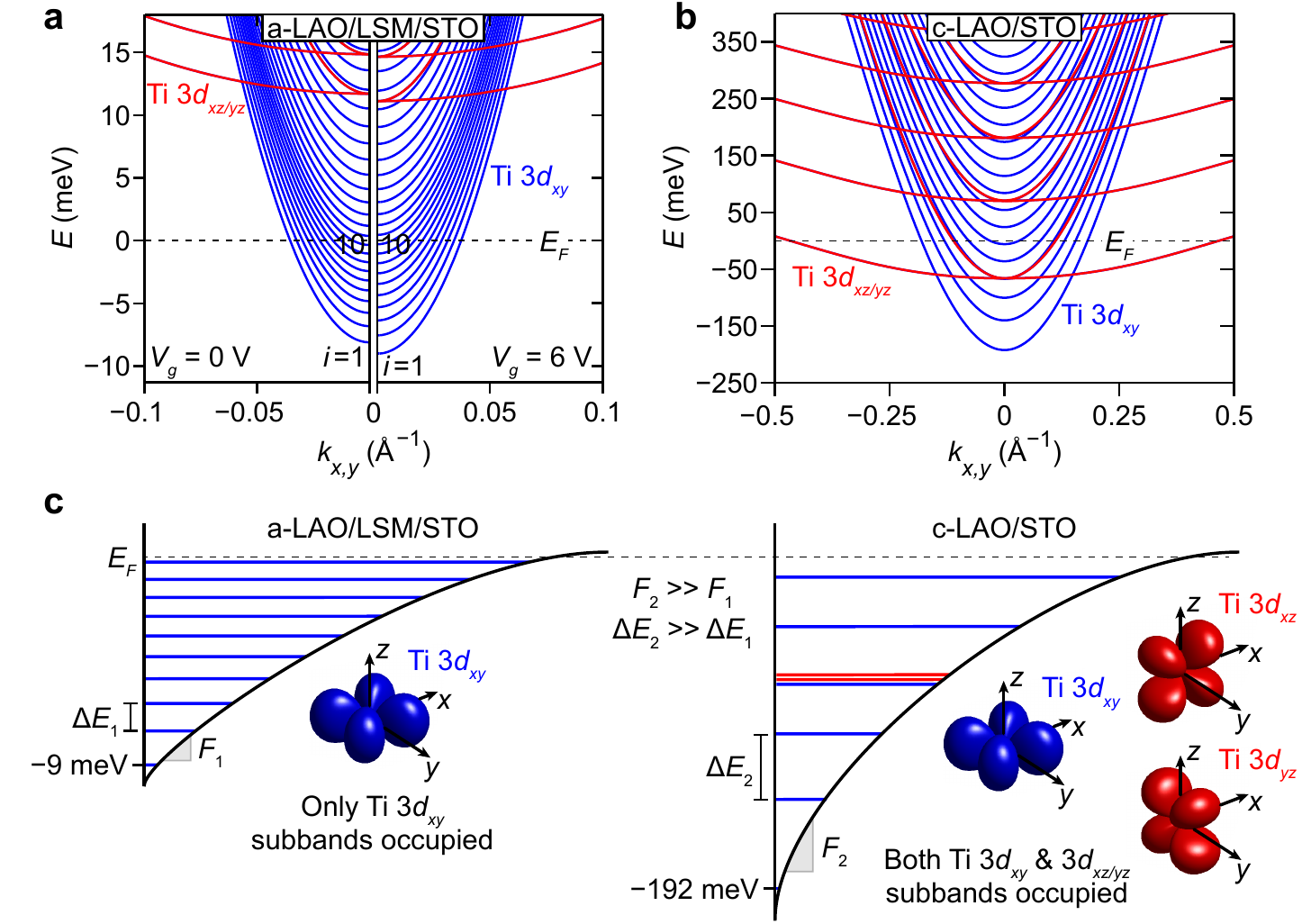}
	\caption{a, DFT-TB calculations of the modulation-doped oxide interface, a-LAO/LSM/STO. b, DFT-TB calculation of the crystalline-LAO/STO (c-LAO/STO) interface. c, The a-LAO/LSM/STO interface is characterized by having a shallow quantum well and only Ti 3\textit{d$_{xy}$} sub-bands occupied as opposed to the c-LAO/STO interface that holds a very deep quantum well with both Ti 3\textit{d$_{xy}$} and 3\textit{d$_{xz/yz}$} sub-bands occupied.}
	\label{figure5}
\end{figure}

To further understand the nature of the parallel conducting channels, DFT-TB calculations are performed considering a wedge-shaped triangular quantum well (SM, Sec. S3). Figure 5a shows the obtained charge carrier dispersions for \textit{V$_g$} = 0 and 6 V (left and right panels, respectively). These calculations support, for both \textit{V$_g$} = 0 and 6 V, the existence of ten occupied Ti 3\textit{d$_{xy}$} sub-bands that have similar effective mass, Hall mobility, and quantum mobility, and calculated charge carrier densities that are consistent with the experimental data (see Figs. 4c and d). Therefore, our modulation-doped oxide interface, a single quantum well with multiple occupied Ti 3\textit{d$_{xy}$} sub-bands (see Fig. 5c), is quite different from what is found in the canonical LAO/STO interface, for \textit{n$_{s}^{\mathrm{Hall}}$} $>$ 1.7$\times$10$^{13}$ cm$^{-2}$ where all \textit{t$_{2g}$} levels (both Ti 3\textit{d$_{xy}$} and 3\textit{d$_{xz/yz}$} sub-bands) are usually occupied in the quantum well (see Figs. 5b and 5c). This difference is caused by a much lower carrier density in a-LAO/LSM/STO compared to LAO/STO due to the LSM spacer layer in the former case acting as an electron acceptor \cite{chen_extreme_2015}. With a reduced carrier density of $\sim$10$^{12}$ cm$^{-2}$ in a-LAO/LSM/STO, the slope of the confining potential is consequently much lower, resulting in a shallower well compared to other STO-based systems such as LAO/STO (see Fig. 5c). With the prevailing population of $3d_{xy}$ sub-bands, the Rashba spin-orbital interaction is expected to be much weaker in our modulation-doped interface compared to the LAO/STO interface with orbital-mixing character \cite{fete_large_2014}. Importantly, the multiple parallel conducting two-dimensional sub-bands differentiate complex oxides, particularly our modulation-doped oxide interface, from conventional semiconductor 2-DEGs with the potential of exploring unusual and exotic physics based on STO.

In summary, we present the observation of quantized Hall resistance with integer step size in two-dimensional conducting complex oxide interfaces based on the patterned a-LAO/LSM/STO structure, revealing that the complex oxide interface consists of a single quantum well with multiple parallel Ti 3\textit{d$_{xy}$} sub-bands. Importantly, as the electrons are confined in STO, the multiple-subband physical picture is expected to be a common feature of all STO-based conducting systems. This insight paves the way for demonstrating new physics in devices based on complex oxide interfaces.

\begin{acknowledgments}
The authors gratefully acknowledge the discussions with A. Smith, J. Levy, S. Ilani, J. Folk, S.L. Folk, G.B.S. Khalsa, and M. Schecter, and the technical assistance from J. Geyti, N. Payami, C.B. S{\o}rensen, S. Upadhyay, C. Olsen, and A. Jellinggaard. Funding from the Danish Agency for Science, Technology and Innovation, and the Lundbeck Foundation are acknowledged. The Center for Quantum Devices is supported by the Danish National Research Foundation.
\end{acknowledgments}


\begin{thebibliography}{41}%
	\makeatletter
	\providecommand \@ifxundefined [1]{%
		\@ifx{#1\undefined}
	}%
	\providecommand \@ifnum [1]{%
		\ifnum #1\expandafter \@firstoftwo
		\else \expandafter \@secondoftwo
		\fi
	}%
	\providecommand \@ifx [1]{%
		\ifx #1\expandafter \@firstoftwo
		\else \expandafter \@secondoftwo
		\fi
	}%
	\providecommand \natexlab [1]{#1}%
	\providecommand \enquote  [1]{``#1''}%
	\providecommand \bibnamefont  [1]{#1}%
	\providecommand \bibfnamefont [1]{#1}%
	\providecommand \citenamefont [1]{#1}%
	\providecommand \href@noop [0]{\@secondoftwo}%
	\providecommand \href [0]{\begingroup \@sanitize@url \@href}%
	\providecommand \@href[1]{\@@startlink{#1}\@@href}%
	\providecommand \@@href[1]{\endgroup#1\@@endlink}%
	\providecommand \@sanitize@url [0]{\catcode `\\12\catcode `\$12\catcode
		`\&12\catcode `\#12\catcode `\^12\catcode `\_12\catcode `\%12\relax}%
	\providecommand \@@startlink[1]{}%
	\providecommand \@@endlink[0]{}%
	\providecommand \url  [0]{\begingroup\@sanitize@url \@url }%
	\providecommand \@url [1]{\endgroup\@href {#1}{\urlprefix }}%
	\providecommand \urlprefix  [0]{URL }%
	\providecommand \Eprint [0]{\href }%
	\providecommand \doibase [0]{http://dx.doi.org/}%
	\providecommand \selectlanguage [0]{\@gobble}%
	\providecommand \bibinfo  [0]{\@secondoftwo}%
	\providecommand \bibfield  [0]{\@secondoftwo}%
	\providecommand \translation [1]{[#1]}%
	\providecommand \BibitemOpen [0]{}%
	\providecommand \bibitemStop [0]{}%
	\providecommand \bibitemNoStop [0]{.\EOS\space}%
	\providecommand \EOS [0]{\spacefactor3000\relax}%
	\providecommand \BibitemShut  [1]{\csname bibitem#1\endcsname}%
	\let\auto@bib@innerbib\@empty
	\bibitem [{\citenamefont {Klitzing}\ \emph {et~al.}(1980)\citenamefont
		{Klitzing}, \citenamefont {Dorda},\ and\ \citenamefont
		{Pepper}}]{klitzing_new_1980}%
	\BibitemOpen
	\bibfield  {author} {\bibinfo {author} {\bibfnamefont {K.~v.}\ \bibnamefont
			{Klitzing}}, \bibinfo {author} {\bibfnamefont {G.}~\bibnamefont {Dorda}}, \
		and\ \bibinfo {author} {\bibfnamefont {M.}~\bibnamefont {Pepper}},\ }\href
	{http://journals.aps.org/prl/abstract/10.1103/PhysRevLett.45.494} {\bibfield
		{journal} {\bibinfo  {journal} {Physical Review Letters}\ }\textbf {\bibinfo
			{volume} {45}},\ \bibinfo {pages} {494} (\bibinfo {year} {1980})}\BibitemShut
	{NoStop}%
	\bibitem [{\citenamefont {Tsui}\ \emph {et~al.}(1982)\citenamefont {Tsui},
		\citenamefont {Stormer},\ and\ \citenamefont
		{Gossard}}]{tsui_two-dimensional_1982}%
	\BibitemOpen
	\bibfield  {author} {\bibinfo {author} {\bibfnamefont {D.~C.}\ \bibnamefont
			{Tsui}}, \bibinfo {author} {\bibfnamefont {H.~L.}\ \bibnamefont {Stormer}}, \
		and\ \bibinfo {author} {\bibfnamefont {A.~C.}\ \bibnamefont {Gossard}},\
	}\href {http://journals.aps.org/prl/abstract/10.1103/PhysRevLett.48.1559}
	{\bibfield  {journal} {\bibinfo  {journal} {Physical Review Letters}\
		}\textbf {\bibinfo {volume} {48}},\ \bibinfo {pages} {1559} (\bibinfo {year}
		{1982})}\BibitemShut {NoStop}%
	\bibitem [{\citenamefont {Cao}\ \emph {et~al.}(2012)\citenamefont {Cao},
		\citenamefont {Tian}, \citenamefont {Miotkowski}, \citenamefont {Shen},
		\citenamefont {Hu}, \citenamefont {Qiao},\ and\ \citenamefont
		{Chen}}]{cao_quantized_2012}%
	\BibitemOpen
	\bibfield  {author} {\bibinfo {author} {\bibfnamefont {H.}~\bibnamefont
			{Cao}}, \bibinfo {author} {\bibfnamefont {J.}~\bibnamefont {Tian}}, \bibinfo
		{author} {\bibfnamefont {I.}~\bibnamefont {Miotkowski}}, \bibinfo {author}
		{\bibfnamefont {T.}~\bibnamefont {Shen}}, \bibinfo {author} {\bibfnamefont
			{J.}~\bibnamefont {Hu}}, \bibinfo {author} {\bibfnamefont {S.}~\bibnamefont
			{Qiao}}, \ and\ \bibinfo {author} {\bibfnamefont {Y.~P.}\ \bibnamefont
			{Chen}},\ }\href {\doibase 10.1103/PhysRevLett.108.216803} {\bibfield
		{journal} {\bibinfo  {journal} {Physical Review Letters}\ }\textbf {\bibinfo
			{volume} {108}},\ \bibinfo {pages} {216803} (\bibinfo {year}
		{2012})}\BibitemShut {NoStop}%
	\bibitem [{\citenamefont {Zhang}\ \emph {et~al.}(2005)\citenamefont {Zhang},
		\citenamefont {Tan}, \citenamefont {Stormer},\ and\ \citenamefont
		{Kim}}]{zhang_experimental_2005}%
	\BibitemOpen
	\bibfield  {author} {\bibinfo {author} {\bibfnamefont {Y.}~\bibnamefont
			{Zhang}}, \bibinfo {author} {\bibfnamefont {Y.-W.}\ \bibnamefont {Tan}},
		\bibinfo {author} {\bibfnamefont {H.~L.}\ \bibnamefont {Stormer}}, \ and\
		\bibinfo {author} {\bibfnamefont {P.}~\bibnamefont {Kim}},\ }\href {\doibase
		10.1038/nature04235} {\bibfield  {journal} {\bibinfo  {journal} {Nature}\
		}\textbf {\bibinfo {volume} {438}},\ \bibinfo {pages} {201} (\bibinfo {year}
		{2005})}\BibitemShut {NoStop}%
	\bibitem [{\citenamefont {Tsukazaki}\ \emph {et~al.}(2007)\citenamefont
		{Tsukazaki}, \citenamefont {Ohtomo}, \citenamefont {Kita}, \citenamefont
		{Ohno}, \citenamefont {Ohno},\ and\ \citenamefont
		{Kawasaki}}]{tsukazaki_quantum_2007}%
	\BibitemOpen
	\bibfield  {author} {\bibinfo {author} {\bibfnamefont {A.}~\bibnamefont
			{Tsukazaki}}, \bibinfo {author} {\bibfnamefont {A.}~\bibnamefont {Ohtomo}},
		\bibinfo {author} {\bibfnamefont {T.}~\bibnamefont {Kita}}, \bibinfo {author}
		{\bibfnamefont {Y.}~\bibnamefont {Ohno}}, \bibinfo {author} {\bibfnamefont
			{H.}~\bibnamefont {Ohno}}, \ and\ \bibinfo {author} {\bibfnamefont
			{M.}~\bibnamefont {Kawasaki}},\ }\href {\doibase 10.1126/science.1137430}
	{\bibfield  {journal} {\bibinfo  {journal} {Science}\ }\textbf {\bibinfo
			{volume} {315}},\ \bibinfo {pages} {1388} (\bibinfo {year}
		{2007})}\BibitemShut {NoStop}%
	\bibitem [{\citenamefont {Tsukazaki}\ \emph {et~al.}(2010)\citenamefont
		{Tsukazaki}, \citenamefont {Akasaka}, \citenamefont {Nakahara}, \citenamefont
		{Ohno}, \citenamefont {Ohno}, \citenamefont {Maryenko}, \citenamefont
		{Ohtomo},\ and\ \citenamefont {Kawasaki}}]{tsukazaki_observation_2010}%
	\BibitemOpen
	\bibfield  {author} {\bibinfo {author} {\bibfnamefont {A.}~\bibnamefont
			{Tsukazaki}}, \bibinfo {author} {\bibfnamefont {S.}~\bibnamefont {Akasaka}},
		\bibinfo {author} {\bibfnamefont {K.}~\bibnamefont {Nakahara}}, \bibinfo
		{author} {\bibfnamefont {Y.}~\bibnamefont {Ohno}}, \bibinfo {author}
		{\bibfnamefont {H.}~\bibnamefont {Ohno}}, \bibinfo {author} {\bibfnamefont
			{D.}~\bibnamefont {Maryenko}}, \bibinfo {author} {\bibfnamefont
			{A.}~\bibnamefont {Ohtomo}}, \ and\ \bibinfo {author} {\bibfnamefont
			{M.}~\bibnamefont {Kawasaki}},\ }\href {\doibase 10.1038/nmat2874} {\bibfield
		{journal} {\bibinfo  {journal} {Nature Materials}\ }\textbf {\bibinfo
			{volume} {9}},\ \bibinfo {pages} {889} (\bibinfo {year} {2010})}\BibitemShut
	{NoStop}%
	\bibitem [{\citenamefont {Ohtomo}\ and\ \citenamefont
		{Hwang}(2004)}]{ohtomo_high-mobility_2004}%
	\BibitemOpen
	\bibfield  {author} {\bibinfo {author} {\bibfnamefont {A.}~\bibnamefont
			{Ohtomo}}\ and\ \bibinfo {author} {\bibfnamefont {H.~Y.}\ \bibnamefont
			{Hwang}},\ }\href
	{http://www.nature.com/nature/journal/v427/n6973/abs/nature02308.html}
	{\bibfield  {journal} {\bibinfo  {journal} {Nature}\ }\textbf {\bibinfo
			{volume} {427}},\ \bibinfo {pages} {423} (\bibinfo {year}
		{2004})}\BibitemShut {NoStop}%
	\bibitem [{\citenamefont {Chen}\ \emph {et~al.}(2013)\citenamefont {Chen},
		\citenamefont {Bovet}, \citenamefont {Trier}, \citenamefont {Christensen},
		\citenamefont {Qu}, \citenamefont {Andersen}, \citenamefont {Kasama},
		\citenamefont {Zhang}, \citenamefont {Giraud}, \citenamefont {Dufouleur},
		\citenamefont {Jespersen}, \citenamefont {Sun}, \citenamefont {Smith},
		\citenamefont {Nygård}, \citenamefont {Lu}, \citenamefont {Büchner},
		\citenamefont {Shen}, \citenamefont {Linderoth},\ and\ \citenamefont
		{Pryds}}]{chen_high-mobility_2013}%
	\BibitemOpen
	\bibfield  {author} {\bibinfo {author} {\bibfnamefont {Y.~Z.}\ \bibnamefont
			{Chen}}, \bibinfo {author} {\bibfnamefont {N.}~\bibnamefont {Bovet}},
		\bibinfo {author} {\bibfnamefont {F.}~\bibnamefont {Trier}}, \bibinfo
		{author} {\bibfnamefont {D.~V.}\ \bibnamefont {Christensen}}, \bibinfo
		{author} {\bibfnamefont {F.~M.}\ \bibnamefont {Qu}}, \bibinfo {author}
		{\bibfnamefont {N.~H.}\ \bibnamefont {Andersen}}, \bibinfo {author}
		{\bibfnamefont {T.}~\bibnamefont {Kasama}}, \bibinfo {author} {\bibfnamefont
			{W.}~\bibnamefont {Zhang}}, \bibinfo {author} {\bibfnamefont
			{R.}~\bibnamefont {Giraud}}, \bibinfo {author} {\bibfnamefont
			{J.}~\bibnamefont {Dufouleur}}, \bibinfo {author} {\bibfnamefont {T.~S.}\
			\bibnamefont {Jespersen}}, \bibinfo {author} {\bibfnamefont {J.~R.}\
			\bibnamefont {Sun}}, \bibinfo {author} {\bibfnamefont {A.}~\bibnamefont
			{Smith}}, \bibinfo {author} {\bibfnamefont {J.}~\bibnamefont {Nygård}},
		\bibinfo {author} {\bibfnamefont {L.}~\bibnamefont {Lu}}, \bibinfo {author}
		{\bibfnamefont {B.}~\bibnamefont {Büchner}}, \bibinfo {author}
		{\bibfnamefont {B.~G.}\ \bibnamefont {Shen}}, \bibinfo {author}
		{\bibfnamefont {S.}~\bibnamefont {Linderoth}}, \ and\ \bibinfo {author}
		{\bibfnamefont {N.}~\bibnamefont {Pryds}},\ }\href {\doibase
		10.1038/ncomms2394} {\bibfield  {journal} {\bibinfo  {journal} {Nature
				Communications}\ }\textbf {\bibinfo {volume} {4}},\ \bibinfo {pages} {1371}
		(\bibinfo {year} {2013})}\BibitemShut {NoStop}%
	\bibitem [{\citenamefont {Falson}\ \emph {et~al.}(2015)\citenamefont {Falson},
		\citenamefont {Maryenko}, \citenamefont {Friess}, \citenamefont {Zhang},
		\citenamefont {Kozuka}, \citenamefont {Tsukazaki}, \citenamefont {Smet},\
		and\ \citenamefont {Kawasaki}}]{falson_even-denominator_2015}%
	\BibitemOpen
	\bibfield  {author} {\bibinfo {author} {\bibfnamefont {J.}~\bibnamefont
			{Falson}}, \bibinfo {author} {\bibfnamefont {D.}~\bibnamefont {Maryenko}},
		\bibinfo {author} {\bibfnamefont {B.}~\bibnamefont {Friess}}, \bibinfo
		{author} {\bibfnamefont {D.}~\bibnamefont {Zhang}}, \bibinfo {author}
		{\bibfnamefont {Y.}~\bibnamefont {Kozuka}}, \bibinfo {author} {\bibfnamefont
			{A.}~\bibnamefont {Tsukazaki}}, \bibinfo {author} {\bibfnamefont {J.~H.}\
			\bibnamefont {Smet}}, \ and\ \bibinfo {author} {\bibfnamefont
			{M.}~\bibnamefont {Kawasaki}},\ }\href {\doibase 10.1038/nphys3259}
	{\bibfield  {journal} {\bibinfo  {journal} {Nature Physics}\ }\textbf
		{\bibinfo {volume} {11}},\ \bibinfo {pages} {347} (\bibinfo {year}
		{2015})}\BibitemShut {NoStop}%
	\bibitem [{\citenamefont {Thiel}\ \emph {et~al.}(2006)\citenamefont {Thiel},
		\citenamefont {Hammerl}, \citenamefont {Schmehl}, \citenamefont {Schneider},\
		and\ \citenamefont {Mannhart}}]{thiel_tunable_2006}%
	\BibitemOpen
	\bibfield  {author} {\bibinfo {author} {\bibfnamefont {S.}~\bibnamefont
			{Thiel}}, \bibinfo {author} {\bibfnamefont {G.}~\bibnamefont {Hammerl}},
		\bibinfo {author} {\bibfnamefont {A.}~\bibnamefont {Schmehl}}, \bibinfo
		{author} {\bibfnamefont {C.~W.}\ \bibnamefont {Schneider}}, \ and\ \bibinfo
		{author} {\bibfnamefont {J.}~\bibnamefont {Mannhart}},\ }\href {\doibase
		10.1126/science.1131091} {\bibfield  {journal} {\bibinfo  {journal}
			{Science}\ }\textbf {\bibinfo {volume} {313}},\ \bibinfo {pages} {1942}
		(\bibinfo {year} {2006})}\BibitemShut {NoStop}%
	\bibitem [{\citenamefont {Christensen}\ \emph {et~al.}(2013)\citenamefont
		{Christensen}, \citenamefont {Trier}, \citenamefont {Chen}, \citenamefont
		{Smith}, \citenamefont {Nygård},\ and\ \citenamefont
		{Pryds}}]{christensen_controlling_2013}%
	\BibitemOpen
	\bibfield  {author} {\bibinfo {author} {\bibfnamefont {D.~V.}\ \bibnamefont
			{Christensen}}, \bibinfo {author} {\bibfnamefont {F.}~\bibnamefont {Trier}},
		\bibinfo {author} {\bibfnamefont {Y.~Z.}\ \bibnamefont {Chen}}, \bibinfo
		{author} {\bibfnamefont {A.}~\bibnamefont {Smith}}, \bibinfo {author}
		{\bibfnamefont {J.}~\bibnamefont {Nygård}}, \ and\ \bibinfo {author}
		{\bibfnamefont {N.}~\bibnamefont {Pryds}},\ }\href {\doibase
		10.1063/1.4775669} {\bibfield  {journal} {\bibinfo  {journal} {Applied
				Physics Letters}\ }\textbf {\bibinfo {volume} {102}},\ \bibinfo {pages}
		{021602} (\bibinfo {year} {2013})}\BibitemShut {NoStop}%
	\bibitem [{\citenamefont {Reyren}\ \emph {et~al.}(2007)\citenamefont {Reyren},
		\citenamefont {Thiel}, \citenamefont {Caviglia}, \citenamefont {Kourkoutis},
		\citenamefont {Hammerl}, \citenamefont {Richter}, \citenamefont {Schneider},
		\citenamefont {Kopp}, \citenamefont {Ruetschi}, \citenamefont {Jaccard},
		\citenamefont {Gabay}, \citenamefont {Muller}, \citenamefont {Triscone},\
		and\ \citenamefont {Mannhart}}]{reyren_superconducting_2007}%
	\BibitemOpen
	\bibfield  {author} {\bibinfo {author} {\bibfnamefont {N.}~\bibnamefont
			{Reyren}}, \bibinfo {author} {\bibfnamefont {S.}~\bibnamefont {Thiel}},
		\bibinfo {author} {\bibfnamefont {A.~D.}\ \bibnamefont {Caviglia}}, \bibinfo
		{author} {\bibfnamefont {L.~F.}\ \bibnamefont {Kourkoutis}}, \bibinfo
		{author} {\bibfnamefont {G.}~\bibnamefont {Hammerl}}, \bibinfo {author}
		{\bibfnamefont {C.}~\bibnamefont {Richter}}, \bibinfo {author} {\bibfnamefont
			{C.~W.}\ \bibnamefont {Schneider}}, \bibinfo {author} {\bibfnamefont
			{T.}~\bibnamefont {Kopp}}, \bibinfo {author} {\bibfnamefont {A.-S.}\
			\bibnamefont {Ruetschi}}, \bibinfo {author} {\bibfnamefont {D.}~\bibnamefont
			{Jaccard}}, \bibinfo {author} {\bibfnamefont {M.}~\bibnamefont {Gabay}},
		\bibinfo {author} {\bibfnamefont {D.~A.}\ \bibnamefont {Muller}}, \bibinfo
		{author} {\bibfnamefont {J.-M.}\ \bibnamefont {Triscone}}, \ and\ \bibinfo
		{author} {\bibfnamefont {J.}~\bibnamefont {Mannhart}},\ }\href {\doibase
		10.1126/science.1146006} {\bibfield  {journal} {\bibinfo  {journal}
			{Science}\ }\textbf {\bibinfo {volume} {317}},\ \bibinfo {pages} {1196}
		(\bibinfo {year} {2007})}\BibitemShut {NoStop}%
	\bibitem [{\citenamefont {Caviglia}\ \emph {et~al.}(2008)\citenamefont
		{Caviglia}, \citenamefont {Gariglio}, \citenamefont {Reyren}, \citenamefont
		{Jaccard}, \citenamefont {Schneider}, \citenamefont {Gabay}, \citenamefont
		{Thiel}, \citenamefont {Hammerl}, \citenamefont {Mannhart},\ and\
		\citenamefont {Triscone}}]{caviglia_electric_2008}%
	\BibitemOpen
	\bibfield  {author} {\bibinfo {author} {\bibfnamefont {A.~D.}\ \bibnamefont
			{Caviglia}}, \bibinfo {author} {\bibfnamefont {S.}~\bibnamefont {Gariglio}},
		\bibinfo {author} {\bibfnamefont {N.}~\bibnamefont {Reyren}}, \bibinfo
		{author} {\bibfnamefont {D.}~\bibnamefont {Jaccard}}, \bibinfo {author}
		{\bibfnamefont {T.}~\bibnamefont {Schneider}}, \bibinfo {author}
		{\bibfnamefont {M.}~\bibnamefont {Gabay}}, \bibinfo {author} {\bibfnamefont
			{S.}~\bibnamefont {Thiel}}, \bibinfo {author} {\bibfnamefont
			{G.}~\bibnamefont {Hammerl}}, \bibinfo {author} {\bibfnamefont
			{J.}~\bibnamefont {Mannhart}}, \ and\ \bibinfo {author} {\bibfnamefont
			{J.-M.}\ \bibnamefont {Triscone}},\ }\href {\doibase 10.1038/nature07576}
	{\bibfield  {journal} {\bibinfo  {journal} {Nature}\ }\textbf {\bibinfo
			{volume} {456}},\ \bibinfo {pages} {624} (\bibinfo {year}
		{2008})}\BibitemShut {NoStop}%
	\bibitem [{\citenamefont {Brinkman}\ \emph {et~al.}(2007)\citenamefont
		{Brinkman}, \citenamefont {Huijben}, \citenamefont {van Zalk}, \citenamefont
		{Huijben}, \citenamefont {Zeitler}, \citenamefont {Maan}, \citenamefont
		{van~der Wiel}, \citenamefont {Rijnders}, \citenamefont {Blank},\ and\
		\citenamefont {Hilgenkamp}}]{brinkman_magnetic_2007}%
	\BibitemOpen
	\bibfield  {author} {\bibinfo {author} {\bibfnamefont {A.}~\bibnamefont
			{Brinkman}}, \bibinfo {author} {\bibfnamefont {M.}~\bibnamefont {Huijben}},
		\bibinfo {author} {\bibfnamefont {M.}~\bibnamefont {van Zalk}}, \bibinfo
		{author} {\bibfnamefont {J.}~\bibnamefont {Huijben}}, \bibinfo {author}
		{\bibfnamefont {U.}~\bibnamefont {Zeitler}}, \bibinfo {author} {\bibfnamefont
			{J.~C.}\ \bibnamefont {Maan}}, \bibinfo {author} {\bibfnamefont {W.~G.}\
			\bibnamefont {van~der Wiel}}, \bibinfo {author} {\bibfnamefont
			{G.}~\bibnamefont {Rijnders}}, \bibinfo {author} {\bibfnamefont {D.~H.~A.}\
			\bibnamefont {Blank}}, \ and\ \bibinfo {author} {\bibfnamefont
			{H.}~\bibnamefont {Hilgenkamp}},\ }\href {\doibase 10.1038/nmat1931}
	{\bibfield  {journal} {\bibinfo  {journal} {Nature Materials}\ }\textbf
		{\bibinfo {volume} {6}},\ \bibinfo {pages} {493} (\bibinfo {year}
		{2007})}\BibitemShut {NoStop}%
	\bibitem [{\citenamefont {Sulpizio}\ \emph {et~al.}(2014)\citenamefont
		{Sulpizio}, \citenamefont {Ilani}, \citenamefont {Irvin},\ and\ \citenamefont
		{Levy}}]{sulpizio_nanoscale_2014}%
	\BibitemOpen
	\bibfield  {author} {\bibinfo {author} {\bibfnamefont {J.~A.}\ \bibnamefont
			{Sulpizio}}, \bibinfo {author} {\bibfnamefont {S.}~\bibnamefont {Ilani}},
		\bibinfo {author} {\bibfnamefont {P.}~\bibnamefont {Irvin}}, \ and\ \bibinfo
		{author} {\bibfnamefont {J.}~\bibnamefont {Levy}},\ }\href {\doibase
		10.1146/annurev-matsci-070813-113437} {\bibfield  {journal} {\bibinfo
			{journal} {Annual Review of Materials Research}\ }\textbf {\bibinfo {volume}
			{44}},\ \bibinfo {pages} {117} (\bibinfo {year} {2014})}\BibitemShut
	{NoStop}%
	\bibitem [{\citenamefont {Delugas}\ \emph {et~al.}(2011)\citenamefont
		{Delugas}, \citenamefont {Filippetti}, \citenamefont {Fiorentini},
		\citenamefont {Bilc}, \citenamefont {Fontaine},\ and\ \citenamefont
		{Ghosez}}]{delugas_spontaneous_2011}%
	\BibitemOpen
	\bibfield  {author} {\bibinfo {author} {\bibfnamefont {P.}~\bibnamefont
			{Delugas}}, \bibinfo {author} {\bibfnamefont {A.}~\bibnamefont {Filippetti}},
		\bibinfo {author} {\bibfnamefont {V.}~\bibnamefont {Fiorentini}}, \bibinfo
		{author} {\bibfnamefont {D.~I.}\ \bibnamefont {Bilc}}, \bibinfo {author}
		{\bibfnamefont {D.}~\bibnamefont {Fontaine}}, \ and\ \bibinfo {author}
		{\bibfnamefont {P.}~\bibnamefont {Ghosez}},\ }\href {\doibase
		10.1103/PhysRevLett.106.166807} {\bibfield  {journal} {\bibinfo  {journal}
			{Physical Review Letters}\ }\textbf {\bibinfo {volume} {106}},\ \bibinfo
		{pages} {166807} (\bibinfo {year} {2011})}\BibitemShut {NoStop}%
	\bibitem [{\citenamefont {Son}\ \emph {et~al.}(2009)\citenamefont {Son},
		\citenamefont {Cho}, \citenamefont {Lee}, \citenamefont {Lee},\ and\
		\citenamefont {Han}}]{son_density_2009}%
	\BibitemOpen
	\bibfield  {author} {\bibinfo {author} {\bibfnamefont {W.-J.}\ \bibnamefont
			{Son}}, \bibinfo {author} {\bibfnamefont {E.}~\bibnamefont {Cho}}, \bibinfo
		{author} {\bibfnamefont {B.}~\bibnamefont {Lee}}, \bibinfo {author}
		{\bibfnamefont {J.}~\bibnamefont {Lee}}, \ and\ \bibinfo {author}
		{\bibfnamefont {S.}~\bibnamefont {Han}},\ }\href {\doibase
		10.1103/PhysRevB.79.245411} {\bibfield  {journal} {\bibinfo  {journal}
			{Physical Review B}\ }\textbf {\bibinfo {volume} {79}},\ \bibinfo {pages}
		{245411} (\bibinfo {year} {2009})}\BibitemShut {NoStop}%
	\bibitem [{\citenamefont {Khalsa}\ and\ \citenamefont
		{MacDonald}(2012)}]{khalsa_theory_2012}%
	\BibitemOpen
	\bibfield  {author} {\bibinfo {author} {\bibfnamefont {G.}~\bibnamefont
			{Khalsa}}\ and\ \bibinfo {author} {\bibfnamefont {A.~H.}\ \bibnamefont
			{MacDonald}},\ }\href {\doibase 10.1103/PhysRevB.86.125121} {\bibfield
		{journal} {\bibinfo  {journal} {Physical Review B}\ }\textbf {\bibinfo
			{volume} {86}},\ \bibinfo {pages} {125121} (\bibinfo {year}
		{2012})}\BibitemShut {NoStop}%
	\bibitem [{\citenamefont {Breitschaft}\ \emph {et~al.}(2010)\citenamefont
		{Breitschaft}, \citenamefont {Tinkl}, \citenamefont {Pavlenko}, \citenamefont
		{Paetel}, \citenamefont {Richter}, \citenamefont {Kirtley}, \citenamefont
		{Liao}, \citenamefont {Hammerl}, \citenamefont {Eyert}, \citenamefont
		{Kopp},\ and\ \citenamefont {Mannhart}}]{breitschaft_two-dimensional_2010}%
	\BibitemOpen
	\bibfield  {author} {\bibinfo {author} {\bibfnamefont {M.}~\bibnamefont
			{Breitschaft}}, \bibinfo {author} {\bibfnamefont {V.}~\bibnamefont {Tinkl}},
		\bibinfo {author} {\bibfnamefont {N.}~\bibnamefont {Pavlenko}}, \bibinfo
		{author} {\bibfnamefont {S.}~\bibnamefont {Paetel}}, \bibinfo {author}
		{\bibfnamefont {C.}~\bibnamefont {Richter}}, \bibinfo {author} {\bibfnamefont
			{J.~R.}\ \bibnamefont {Kirtley}}, \bibinfo {author} {\bibfnamefont {Y.~C.}\
			\bibnamefont {Liao}}, \bibinfo {author} {\bibfnamefont {G.}~\bibnamefont
			{Hammerl}}, \bibinfo {author} {\bibfnamefont {V.}~\bibnamefont {Eyert}},
		\bibinfo {author} {\bibfnamefont {T.}~\bibnamefont {Kopp}}, \ and\ \bibinfo
		{author} {\bibfnamefont {J.}~\bibnamefont {Mannhart}},\ }\href {\doibase
		10.1103/PhysRevB.81.153414} {\bibfield  {journal} {\bibinfo  {journal}
			{Physical Review B}\ }\textbf {\bibinfo {volume} {81}},\ \bibinfo {pages}
		{153414} (\bibinfo {year} {2010})}\BibitemShut {NoStop}%
	\bibitem [{\citenamefont {Joshua}\ \emph {et~al.}(2012)\citenamefont {Joshua},
		\citenamefont {Pecker}, \citenamefont {Ruhman}, \citenamefont {Altman},\ and\
		\citenamefont {Ilani}}]{joshua_universal_2012}%
	\BibitemOpen
	\bibfield  {author} {\bibinfo {author} {\bibfnamefont {A.}~\bibnamefont
			{Joshua}}, \bibinfo {author} {\bibfnamefont {S.}~\bibnamefont {Pecker}},
		\bibinfo {author} {\bibfnamefont {J.}~\bibnamefont {Ruhman}}, \bibinfo
		{author} {\bibfnamefont {E.}~\bibnamefont {Altman}}, \ and\ \bibinfo {author}
		{\bibfnamefont {S.}~\bibnamefont {Ilani}},\ }\href {\doibase
		10.1038/ncomms2116} {\bibfield  {journal} {\bibinfo  {journal} {Nature
				Communications}\ }\textbf {\bibinfo {volume} {3}},\ \bibinfo {pages} {1129}
		(\bibinfo {year} {2012})}\BibitemShut {NoStop}%
	\bibitem [{\citenamefont {Gabay}\ and\ \citenamefont
		{Triscone}(2013)}]{gabay_oxide_2013}%
	\BibitemOpen
	\bibfield  {author} {\bibinfo {author} {\bibfnamefont {M.}~\bibnamefont
			{Gabay}}\ and\ \bibinfo {author} {\bibfnamefont {J.-M.}\ \bibnamefont
			{Triscone}},\ }\href
	{http://www.nature.com/nphys/journal/v9/n10/full/nphys2737.html} {\bibfield
		{journal} {\bibinfo  {journal} {Nature Physics}\ }\textbf {\bibinfo {volume}
			{9}},\ \bibinfo {pages} {610} (\bibinfo {year} {2013})}\BibitemShut {NoStop}%
	\bibitem [{\citenamefont {Caviglia}\ \emph
		{et~al.}(2010{\natexlab{a}})\citenamefont {Caviglia}, \citenamefont
		{Gariglio}, \citenamefont {Cancellieri}, \citenamefont {Sac\'{e}p\'{e}},
		\citenamefont {F\^ete}, \citenamefont {Reyren}, \citenamefont {Gabay},
		\citenamefont {Morpurgo},\ and\ \citenamefont
		{Triscone}}]{caviglia_two-dimensional_2010}%
	\BibitemOpen
	\bibfield  {author} {\bibinfo {author} {\bibfnamefont {A.~D.}\ \bibnamefont
			{Caviglia}}, \bibinfo {author} {\bibfnamefont {S.}~\bibnamefont {Gariglio}},
		\bibinfo {author} {\bibfnamefont {C.}~\bibnamefont {Cancellieri}}, \bibinfo
		{author} {\bibfnamefont {B.}~\bibnamefont {Sac\'{e}p\'{e}}}, \bibinfo
		{author} {\bibfnamefont {A.}~\bibnamefont {F\^ete}}, \bibinfo {author}
		{\bibfnamefont {N.}~\bibnamefont {Reyren}}, \bibinfo {author} {\bibfnamefont
			{M.}~\bibnamefont {Gabay}}, \bibinfo {author} {\bibfnamefont {A.~F.}\
			\bibnamefont {Morpurgo}}, \ and\ \bibinfo {author} {\bibfnamefont {J.-M.}\
			\bibnamefont {Triscone}},\ }\href {\doibase 10.1103/PhysRevLett.105.236802}
	{\bibfield  {journal} {\bibinfo  {journal} {Physical Review Letters}\
		}\textbf {\bibinfo {volume} {105}},\ \bibinfo {pages} {236802} (\bibinfo
		{year} {2010}{\natexlab{a}})}\BibitemShut {NoStop}%
	\bibitem [{\citenamefont {Ben~Shalom}\ \emph {et~al.}(2010)\citenamefont
		{Ben~Shalom}, \citenamefont {Ron}, \citenamefont {Palevski},\ and\
		\citenamefont {Dagan}}]{ben_shalom_shubnikovhaas_2010}%
	\BibitemOpen
	\bibfield  {author} {\bibinfo {author} {\bibfnamefont {M.}~\bibnamefont
			{Ben~Shalom}}, \bibinfo {author} {\bibfnamefont {A.}~\bibnamefont {Ron}},
		\bibinfo {author} {\bibfnamefont {A.}~\bibnamefont {Palevski}}, \ and\
		\bibinfo {author} {\bibfnamefont {Y.}~\bibnamefont {Dagan}},\ }\href
	{\doibase 10.1103/PhysRevLett.105.206401} {\bibfield  {journal} {\bibinfo
			{journal} {Physical Review Letters}\ }\textbf {\bibinfo {volume} {105}},\
		\bibinfo {pages} {206401} (\bibinfo {year} {2010})}\BibitemShut {NoStop}%
	\bibitem [{\citenamefont {Xie}\ \emph {et~al.}(2014)\citenamefont {Xie},
		\citenamefont {Bell}, \citenamefont {Kim}, \citenamefont {Inoue},
		\citenamefont {Hikita},\ and\ \citenamefont {Hwang}}]{xie_quantum_2014}%
	\BibitemOpen
	\bibfield  {author} {\bibinfo {author} {\bibfnamefont {Y.}~\bibnamefont
			{Xie}}, \bibinfo {author} {\bibfnamefont {C.}~\bibnamefont {Bell}}, \bibinfo
		{author} {\bibfnamefont {M.}~\bibnamefont {Kim}}, \bibinfo {author}
		{\bibfnamefont {H.}~\bibnamefont {Inoue}}, \bibinfo {author} {\bibfnamefont
			{Y.}~\bibnamefont {Hikita}}, \ and\ \bibinfo {author} {\bibfnamefont {H.~Y.}\
			\bibnamefont {Hwang}},\ }\href {\doibase 10.1016/j.ssc.2014.08.006}
	{\bibfield  {journal} {\bibinfo  {journal} {Solid State Communications}\
		}\textbf {\bibinfo {volume} {197}},\ \bibinfo {pages} {25} (\bibinfo {year}
		{2014})}\BibitemShut {NoStop}%
	\bibitem [{\citenamefont {McCollam}\ \emph {et~al.}(2014)\citenamefont
		{McCollam}, \citenamefont {Wenderich}, \citenamefont {Kruize}, \citenamefont
		{Guduru}, \citenamefont {Molegraaf}, \citenamefont {Huijben}, \citenamefont
		{Koster}, \citenamefont {Blank}, \citenamefont {Rijnders}, \citenamefont
		{Brinkman}, \citenamefont {Hilgenkamp}, \citenamefont {Zeitler},\ and\
		\citenamefont {Maan}}]{mccollam_quantum_2014}%
	\BibitemOpen
	\bibfield  {author} {\bibinfo {author} {\bibfnamefont {A.}~\bibnamefont
			{McCollam}}, \bibinfo {author} {\bibfnamefont {S.}~\bibnamefont {Wenderich}},
		\bibinfo {author} {\bibfnamefont {M.~K.}\ \bibnamefont {Kruize}}, \bibinfo
		{author} {\bibfnamefont {V.~K.}\ \bibnamefont {Guduru}}, \bibinfo {author}
		{\bibfnamefont {H.~J.~A.}\ \bibnamefont {Molegraaf}}, \bibinfo {author}
		{\bibfnamefont {M.}~\bibnamefont {Huijben}}, \bibinfo {author} {\bibfnamefont
			{G.}~\bibnamefont {Koster}}, \bibinfo {author} {\bibfnamefont {D.~H.~A.}\
			\bibnamefont {Blank}}, \bibinfo {author} {\bibfnamefont {G.}~\bibnamefont
			{Rijnders}}, \bibinfo {author} {\bibfnamefont {A.}~\bibnamefont {Brinkman}},
		\bibinfo {author} {\bibfnamefont {H.}~\bibnamefont {Hilgenkamp}}, \bibinfo
		{author} {\bibfnamefont {U.}~\bibnamefont {Zeitler}}, \ and\ \bibinfo
		{author} {\bibfnamefont {J.~C.}\ \bibnamefont {Maan}},\ }\href {\doibase
		10.1063/1.4863786} {\bibfield  {journal} {\bibinfo  {journal} {APL
				Materials}\ }\textbf {\bibinfo {volume} {2}},\ \bibinfo {pages} {022102}
		(\bibinfo {year} {2014})}\BibitemShut {NoStop}%
	\bibitem [{\citenamefont {Son}\ \emph {et~al.}(2010)\citenamefont {Son},
		\citenamefont {Moetakef}, \citenamefont {Jalan}, \citenamefont {Bierwagen},
		\citenamefont {Wright}, \citenamefont {Engel-Herbert},\ and\ \citenamefont
		{Stemmer}}]{son_epitaxial_2010}%
	\BibitemOpen
	\bibfield  {author} {\bibinfo {author} {\bibfnamefont {J.}~\bibnamefont
			{Son}}, \bibinfo {author} {\bibfnamefont {P.}~\bibnamefont {Moetakef}},
		\bibinfo {author} {\bibfnamefont {B.}~\bibnamefont {Jalan}}, \bibinfo
		{author} {\bibfnamefont {O.}~\bibnamefont {Bierwagen}}, \bibinfo {author}
		{\bibfnamefont {N.~J.}\ \bibnamefont {Wright}}, \bibinfo {author}
		{\bibfnamefont {R.}~\bibnamefont {Engel-Herbert}}, \ and\ \bibinfo {author}
		{\bibfnamefont {S.}~\bibnamefont {Stemmer}},\ }\href {\doibase
		10.1038/nmat2750} {\bibfield  {journal} {\bibinfo  {journal} {Nature
				Materials}\ }\textbf {\bibinfo {volume} {9}},\ \bibinfo {pages} {482}
		(\bibinfo {year} {2010})}\BibitemShut {NoStop}%
	\bibitem [{\citenamefont {Chen}\ \emph {et~al.}(2015)\citenamefont {Chen},
		\citenamefont {Trier}, \citenamefont {Wijnands}, \citenamefont {Green},
		\citenamefont {Gauquelin}, \citenamefont {Egoavil}, \citenamefont
		{Christensen}, \citenamefont {Koster}, \citenamefont {Huijben}, \citenamefont
		{Bovet}, \citenamefont {Macke}, \citenamefont {He}, \citenamefont {Sutarto},
		\citenamefont {Andersen}, \citenamefont {Sulpizio}, \citenamefont {Honig},
		\citenamefont {Prawiroatmodjo}, \citenamefont {Jespersen}, \citenamefont
		{Linderoth}, \citenamefont {Ilani}, \citenamefont {Verbeeck}, \citenamefont
		{Van Tendeloo}, \citenamefont {Rijnders}, \citenamefont {Sawatzky},\ and\
		\citenamefont {Pryds}}]{chen_extreme_2015}%
	\BibitemOpen
	\bibfield  {author} {\bibinfo {author} {\bibfnamefont {Y.~Z.}\ \bibnamefont
			{Chen}}, \bibinfo {author} {\bibfnamefont {F.}~\bibnamefont {Trier}},
		\bibinfo {author} {\bibfnamefont {T.}~\bibnamefont {Wijnands}}, \bibinfo
		{author} {\bibfnamefont {R.~J.}\ \bibnamefont {Green}}, \bibinfo {author}
		{\bibfnamefont {N.}~\bibnamefont {Gauquelin}}, \bibinfo {author}
		{\bibfnamefont {R.}~\bibnamefont {Egoavil}}, \bibinfo {author} {\bibfnamefont
			{D.~V.}\ \bibnamefont {Christensen}}, \bibinfo {author} {\bibfnamefont
			{G.}~\bibnamefont {Koster}}, \bibinfo {author} {\bibfnamefont
			{M.}~\bibnamefont {Huijben}}, \bibinfo {author} {\bibfnamefont
			{N.}~\bibnamefont {Bovet}}, \bibinfo {author} {\bibfnamefont
			{S.}~\bibnamefont {Macke}}, \bibinfo {author} {\bibfnamefont
			{F.}~\bibnamefont {He}}, \bibinfo {author} {\bibfnamefont {R.}~\bibnamefont
			{Sutarto}}, \bibinfo {author} {\bibfnamefont {N.~H.}\ \bibnamefont
			{Andersen}}, \bibinfo {author} {\bibfnamefont {J.~A.}\ \bibnamefont
			{Sulpizio}}, \bibinfo {author} {\bibfnamefont {M.}~\bibnamefont {Honig}},
		\bibinfo {author} {\bibfnamefont {G.~E. D.~K.}\ \bibnamefont
			{Prawiroatmodjo}}, \bibinfo {author} {\bibfnamefont {T.~S.}\ \bibnamefont
			{Jespersen}}, \bibinfo {author} {\bibfnamefont {S.}~\bibnamefont
			{Linderoth}}, \bibinfo {author} {\bibfnamefont {S.}~\bibnamefont {Ilani}},
		\bibinfo {author} {\bibfnamefont {J.}~\bibnamefont {Verbeeck}}, \bibinfo
		{author} {\bibfnamefont {G.}~\bibnamefont {Van Tendeloo}}, \bibinfo {author}
		{\bibfnamefont {G.}~\bibnamefont {Rijnders}}, \bibinfo {author}
		{\bibfnamefont {G.~A.}\ \bibnamefont {Sawatzky}}, \ and\ \bibinfo {author}
		{\bibfnamefont {N.}~\bibnamefont {Pryds}},\ }\href {\doibase
		10.1038/nmat4303} {\bibfield  {journal} {\bibinfo  {journal} {Nature
				Materials}\ }\textbf {\bibinfo {volume} {14}},\ \bibinfo {pages} {801}
		(\bibinfo {year} {2015})}\BibitemShut {NoStop}%
	\bibitem [{\citenamefont {Trier}\ \emph {et~al.}(2015)\citenamefont {Trier},
		\citenamefont {Prawiroatmodjo}, \citenamefont {von Soosten}, \citenamefont
		{Christensen}, \citenamefont {Jespersen}, \citenamefont {Chen},\ and\
		\citenamefont {Pryds}}]{trier_patterning_2015}%
	\BibitemOpen
	\bibfield  {author} {\bibinfo {author} {\bibfnamefont {F.}~\bibnamefont
			{Trier}}, \bibinfo {author} {\bibfnamefont {G.~E. D.~K.}\ \bibnamefont
			{Prawiroatmodjo}}, \bibinfo {author} {\bibfnamefont {M.}~\bibnamefont {von
				Soosten}}, \bibinfo {author} {\bibfnamefont {D.~V.}\ \bibnamefont
			{Christensen}}, \bibinfo {author} {\bibfnamefont {T.~S.}\ \bibnamefont
			{Jespersen}}, \bibinfo {author} {\bibfnamefont {Y.~Z.}\ \bibnamefont {Chen}},
		\ and\ \bibinfo {author} {\bibfnamefont {N.}~\bibnamefont {Pryds}},\ }\href
	{\doibase 10.1063/1.4935553} {\bibfield  {journal} {\bibinfo  {journal}
			{Applied Physics Letters}\ }\textbf {\bibinfo {volume} {107}},\ \bibinfo
		{pages} {191604} (\bibinfo {year} {2015})}\BibitemShut {NoStop}%
	\bibitem [{\citenamefont {Schneider}\ \emph {et~al.}(2006)\citenamefont
		{Schneider}, \citenamefont {Thiel}, \citenamefont {Hammerl}, \citenamefont
		{Richter},\ and\ \citenamefont {Mannhart}}]{schneider_microlithography_2006}%
	\BibitemOpen
	\bibfield  {author} {\bibinfo {author} {\bibfnamefont {C.~W.}\ \bibnamefont
			{Schneider}}, \bibinfo {author} {\bibfnamefont {S.}~\bibnamefont {Thiel}},
		\bibinfo {author} {\bibfnamefont {G.}~\bibnamefont {Hammerl}}, \bibinfo
		{author} {\bibfnamefont {C.}~\bibnamefont {Richter}}, \ and\ \bibinfo
		{author} {\bibfnamefont {J.}~\bibnamefont {Mannhart}},\ }\href {\doibase
		10.1063/1.2354422} {\bibfield  {journal} {\bibinfo  {journal} {Applied
				Physics Letters}\ }\textbf {\bibinfo {volume} {89}},\ \bibinfo {pages}
		{122101} (\bibinfo {year} {2006})}\BibitemShut {NoStop}%
	\bibitem [{\citenamefont {Coleridge}\ \emph {et~al.}(1989)\citenamefont
		{Coleridge}, \citenamefont {Stoner},\ and\ \citenamefont
		{Fletcher}}]{coleridge_low-field_1989}%
	\BibitemOpen
	\bibfield  {author} {\bibinfo {author} {\bibfnamefont {P.~T.}\ \bibnamefont
			{Coleridge}}, \bibinfo {author} {\bibfnamefont {R.}~\bibnamefont {Stoner}}, \
		and\ \bibinfo {author} {\bibfnamefont {R.}~\bibnamefont {Fletcher}},\ }\href
	{http://prb.aps.org/abstract/PRB/v39/i2/p1120_1} {\bibfield  {journal}
		{\bibinfo  {journal} {Physical Review B}\ }\textbf {\bibinfo {volume} {39}},\
		\bibinfo {pages} {1120} (\bibinfo {year} {1989})}\BibitemShut {NoStop}%
	\bibitem [{\citenamefont {Zhong}\ \emph {et~al.}(2013)\citenamefont {Zhong},
		\citenamefont {Zhang},\ and\ \citenamefont {Held}}]{zhong_quantum_2013}%
	\BibitemOpen
	\bibfield  {author} {\bibinfo {author} {\bibfnamefont {Z.}~\bibnamefont
			{Zhong}}, \bibinfo {author} {\bibfnamefont {Q.}~\bibnamefont {Zhang}}, \ and\
		\bibinfo {author} {\bibfnamefont {K.}~\bibnamefont {Held}},\ }\href {\doibase
		10.1103/PhysRevB.88.125401} {\bibfield  {journal} {\bibinfo  {journal}
			{Physical Review B}\ }\textbf {\bibinfo {volume} {88}},\ \bibinfo {pages}
		{125401} (\bibinfo {year} {2013})}\BibitemShut {NoStop}%
	\bibitem [{\citenamefont {Stornaiuolo}\ \emph {et~al.}(2012)\citenamefont
		{Stornaiuolo}, \citenamefont {Gariglio}, \citenamefont {Couto}, \citenamefont
		{Fête}, \citenamefont {Caviglia}, \citenamefont {Seyfarth}, \citenamefont
		{Jaccard}, \citenamefont {Morpurgo},\ and\ \citenamefont
		{Triscone}}]{stornaiuolo_-plane_2012}%
	\BibitemOpen
	\bibfield  {author} {\bibinfo {author} {\bibfnamefont {D.}~\bibnamefont
			{Stornaiuolo}}, \bibinfo {author} {\bibfnamefont {S.}~\bibnamefont
			{Gariglio}}, \bibinfo {author} {\bibfnamefont {N.~J.~G.}\ \bibnamefont
			{Couto}}, \bibinfo {author} {\bibfnamefont {A.}~\bibnamefont {Fête}},
		\bibinfo {author} {\bibfnamefont {A.~D.}\ \bibnamefont {Caviglia}}, \bibinfo
		{author} {\bibfnamefont {G.}~\bibnamefont {Seyfarth}}, \bibinfo {author}
		{\bibfnamefont {D.}~\bibnamefont {Jaccard}}, \bibinfo {author} {\bibfnamefont
			{A.~F.}\ \bibnamefont {Morpurgo}}, \ and\ \bibinfo {author} {\bibfnamefont
			{J.-M.}\ \bibnamefont {Triscone}},\ }\href {\doibase 10.1063/1.4768936}
	{\bibfield  {journal} {\bibinfo  {journal} {Applied Physics Letters}\
		}\textbf {\bibinfo {volume} {101}},\ \bibinfo {pages} {222601} (\bibinfo
		{year} {2012})}\BibitemShut {NoStop}%
	\bibitem [{\citenamefont {Ando}\ \emph {et~al.}(1982)\citenamefont {Ando},
		\citenamefont {Fowler},\ and\ \citenamefont {Stern}}]{ando_electronic_1982}%
	\BibitemOpen
	\bibfield  {author} {\bibinfo {author} {\bibfnamefont {T.}~\bibnamefont
			{Ando}}, \bibinfo {author} {\bibfnamefont {A.~B.}\ \bibnamefont {Fowler}}, \
		and\ \bibinfo {author} {\bibfnamefont {F.}~\bibnamefont {Stern}},\ }\href
	{\doibase 10.1103/RevModPhys.54.437} {\bibfield  {journal} {\bibinfo
			{journal} {Rev. Mod. Phys.}\ }\textbf {\bibinfo {volume} {54}},\ \bibinfo
		{pages} {437} (\bibinfo {year} {1982})}\BibitemShut {NoStop}%
	\bibitem [{\citenamefont {Laguta}\ \emph {et~al.}(2002)\citenamefont {Laguta},
		\citenamefont {Glinchuk}, \citenamefont {Kuzian}, \citenamefont {Nokhrin},
		\citenamefont {Bykov}, \citenamefont {Rosa}, \citenamefont {Jastrabik},\ and\
		\citenamefont {Karkut}}]{laguta_photoinduced_2002}%
	\BibitemOpen
	\bibfield  {author} {\bibinfo {author} {\bibfnamefont {V.~V.}\ \bibnamefont
			{Laguta}}, \bibinfo {author} {\bibfnamefont {M.~D.}\ \bibnamefont
			{Glinchuk}}, \bibinfo {author} {\bibfnamefont {R.~O.}\ \bibnamefont
			{Kuzian}}, \bibinfo {author} {\bibfnamefont {S.~N.}\ \bibnamefont {Nokhrin}},
		\bibinfo {author} {\bibfnamefont {I.~P.}\ \bibnamefont {Bykov}}, \bibinfo
		{author} {\bibfnamefont {J.}~\bibnamefont {Rosa}}, \bibinfo {author}
		{\bibfnamefont {L.}~\bibnamefont {Jastrabik}}, \ and\ \bibinfo {author}
		{\bibfnamefont {M.~G.}\ \bibnamefont {Karkut}},\ }\href
	{http://iopscience.iop.org/article/10.1088/0953-8984/14/50/308/meta}
	{\bibfield  {journal} {\bibinfo  {journal} {Journal of Physics: Condensed
				Matter}\ }\textbf {\bibinfo {volume} {14}},\ \bibinfo {pages} {13813}
		(\bibinfo {year} {2002})}\BibitemShut {NoStop}%
	\bibitem [{\citenamefont {Caviglia}\ \emph
		{et~al.}(2010{\natexlab{b}})\citenamefont {Caviglia}, \citenamefont {Gabay},
		\citenamefont {Gariglio}, \citenamefont {Reyren}, \citenamefont
		{Cancellieri},\ and\ \citenamefont {Triscone}}]{caviglia_tunable_2010}%
	\BibitemOpen
	\bibfield  {author} {\bibinfo {author} {\bibfnamefont {A.~D.}\ \bibnamefont
			{Caviglia}}, \bibinfo {author} {\bibfnamefont {M.}~\bibnamefont {Gabay}},
		\bibinfo {author} {\bibfnamefont {S.}~\bibnamefont {Gariglio}}, \bibinfo
		{author} {\bibfnamefont {N.}~\bibnamefont {Reyren}}, \bibinfo {author}
		{\bibfnamefont {C.}~\bibnamefont {Cancellieri}}, \ and\ \bibinfo {author}
		{\bibfnamefont {J.-M.}\ \bibnamefont {Triscone}},\ }\href {\doibase
		10.1103/PhysRevLett.104.126803} {\bibfield  {journal} {\bibinfo  {journal}
			{Physical Review Letters}\ }\textbf {\bibinfo {volume} {104}},\ \bibinfo
		{pages} {126803} (\bibinfo {year} {2010}{\natexlab{b}})}\BibitemShut
	{NoStop}%
	\bibitem [{\citenamefont {van~der Burgt}\ \emph {et~al.}(1995)\citenamefont
		{van~der Burgt}, \citenamefont {Karavolas}, \citenamefont {Peeters},
		\citenamefont {Singleton}, \citenamefont {Nicholas}, \citenamefont {Herlach},
		\citenamefont {Harris}, \citenamefont {Van~Hove},\ and\ \citenamefont
		{Borghs}}]{van_der_burgt_magnetotransport_1995}%
	\BibitemOpen
	\bibfield  {author} {\bibinfo {author} {\bibfnamefont {M.}~\bibnamefont
			{van~der Burgt}}, \bibinfo {author} {\bibfnamefont {V.~C.}\ \bibnamefont
			{Karavolas}}, \bibinfo {author} {\bibfnamefont {F.~M.}\ \bibnamefont
			{Peeters}}, \bibinfo {author} {\bibfnamefont {J.}~\bibnamefont {Singleton}},
		\bibinfo {author} {\bibfnamefont {R.~J.}\ \bibnamefont {Nicholas}}, \bibinfo
		{author} {\bibfnamefont {F.}~\bibnamefont {Herlach}}, \bibinfo {author}
		{\bibfnamefont {J.~J.}\ \bibnamefont {Harris}}, \bibinfo {author}
		{\bibfnamefont {M.}~\bibnamefont {Van~Hove}}, \ and\ \bibinfo {author}
		{\bibfnamefont {G.}~\bibnamefont {Borghs}},\ }\href
	{http://journals.aps.org/prb/abstract/10.1103/PhysRevB.52.12218} {\bibfield
		{journal} {\bibinfo  {journal} {Physical Review B}\ }\textbf {\bibinfo
			{volume} {52}},\ \bibinfo {pages} {12218} (\bibinfo {year}
		{1995})}\BibitemShut {NoStop}%
	\bibitem [{\citenamefont {Grayson}\ and\ \citenamefont
		{Fischer}(2005)}]{grayson_measuring_2005}%
	\BibitemOpen
	\bibfield  {author} {\bibinfo {author} {\bibfnamefont {M.}~\bibnamefont
			{Grayson}}\ and\ \bibinfo {author} {\bibfnamefont {F.}~\bibnamefont
			{Fischer}},\ }\href {\doibase 10.1063/1.1948529} {\bibfield  {journal}
		{\bibinfo  {journal} {Journal of Applied Physics}\ }\textbf {\bibinfo
			{volume} {98}},\ \bibinfo {pages} {013709} (\bibinfo {year}
		{2005})}\BibitemShut {NoStop}%
	\bibitem [{\citenamefont {St\"{o}rmer}\ \emph {et~al.}(1986)\citenamefont
		{St\"{o}rmer}, \citenamefont {Eisenstein}, \citenamefont {Gossard},
		\citenamefont {Wiegmann},\ and\ \citenamefont
		{Baldwin}}]{stormer_quantization_1986}%
	\BibitemOpen
	\bibfield  {author} {\bibinfo {author} {\bibfnamefont {H.~L.}\ \bibnamefont
			{St\"{o}rmer}}, \bibinfo {author} {\bibfnamefont {J.~P.}\ \bibnamefont
			{Eisenstein}}, \bibinfo {author} {\bibfnamefont {A.~C.}\ \bibnamefont
			{Gossard}}, \bibinfo {author} {\bibfnamefont {W.}~\bibnamefont {Wiegmann}}, \
		and\ \bibinfo {author} {\bibfnamefont {K.}~\bibnamefont {Baldwin}},\ }\href
	{\doibase 10.1103/PhysRevLett.56.85} {\bibfield  {journal} {\bibinfo
			{journal} {Phys. Rev. Lett.}\ }\textbf {\bibinfo {volume} {56}},\ \bibinfo
		{pages} {85} (\bibinfo {year} {1986})}\BibitemShut {NoStop}%
	\bibitem [{\citenamefont {Reich}\ \emph {et~al.}(2015)\citenamefont {Reich},
		\citenamefont {Schecter},\ and\ \citenamefont
		{Shklovskii}}]{reich_accumulation_2015}%
	\BibitemOpen
	\bibfield  {author} {\bibinfo {author} {\bibfnamefont {K.~V.}\ \bibnamefont
			{Reich}}, \bibinfo {author} {\bibfnamefont {M.}~\bibnamefont {Schecter}}, \
		and\ \bibinfo {author} {\bibfnamefont {B.~I.}\ \bibnamefont {Shklovskii}},\
	}\href {\doibase 10.1103/PhysRevB.91.115303} {\bibfield  {journal} {\bibinfo
		{journal} {Physical Review B}\ }\textbf {\bibinfo {volume} {91}},\ \bibinfo
	{pages} {115303} (\bibinfo {year} {2015})}\BibitemShut {NoStop}%
\bibitem [{\citenamefont {Pallecchi}\ \emph {et~al.}(2015)\citenamefont
	{Pallecchi}, \citenamefont {Telesio}, \citenamefont {Li}, \citenamefont
	{F\^ete}, \citenamefont {Gariglio}, \citenamefont {Triscone}, \citenamefont
	{Filippetti}, \citenamefont {Delugas}, \citenamefont {Fiorentini},\ and\
	\citenamefont {Marr\'e}}]{pallecchi_giant_2015}%
\BibitemOpen
\bibfield  {author} {\bibinfo {author} {\bibfnamefont {I.}~\bibnamefont
		{Pallecchi}}, \bibinfo {author} {\bibfnamefont {F.}~\bibnamefont {Telesio}},
	\bibinfo {author} {\bibfnamefont {D.}~\bibnamefont {Li}}, \bibinfo {author}
	{\bibfnamefont {A.}~\bibnamefont {F\^ete}}, \bibinfo {author} {\bibfnamefont
		{S.}~\bibnamefont {Gariglio}}, \bibinfo {author} {\bibfnamefont {J.-M.}\
		\bibnamefont {Triscone}}, \bibinfo {author} {\bibfnamefont {A.}~\bibnamefont
		{Filippetti}}, \bibinfo {author} {\bibfnamefont {P.}~\bibnamefont {Delugas}},
	\bibinfo {author} {\bibfnamefont {V.}~\bibnamefont {Fiorentini}}, \ and\
	\bibinfo {author} {\bibfnamefont {D.}~\bibnamefont {Marr\'e}},\ }\href
{\doibase 10.1038/ncomms7678} {\bibfield  {journal} {\bibinfo  {journal}
		{Nature Communications}\ }\textbf {\bibinfo {volume} {6}},\ \bibinfo {pages}
	{6678} (\bibinfo {year} {2015})}\BibitemShut {NoStop}%
\bibitem [{\citenamefont {F\^ete}\ \emph {et~al.}(2014)\citenamefont {F\^ete},
	\citenamefont {Gariglio}, \citenamefont {Berthod}, \citenamefont {Li},
	\citenamefont {Stornaiuolo}, \citenamefont {Gabay},\ and\ \citenamefont
	{Triscone}}]{fete_large_2014}%
\BibitemOpen
\bibfield  {author} {\bibinfo {author} {\bibfnamefont {A.}~\bibnamefont
		{F\^ete}}, \bibinfo {author} {\bibfnamefont {S.}~\bibnamefont {Gariglio}},
	\bibinfo {author} {\bibfnamefont {C.}~\bibnamefont {Berthod}}, \bibinfo
	{author} {\bibfnamefont {D.}~\bibnamefont {Li}}, \bibinfo {author}
	{\bibfnamefont {D.}~\bibnamefont {Stornaiuolo}}, \bibinfo {author}
	{\bibfnamefont {M.}~\bibnamefont {Gabay}}, \ and\ \bibinfo {author}
	{\bibfnamefont {J.-M.}\ \bibnamefont {Triscone}},\ }\href {\doibase
	10.1088/1367-2630/16/11/112002} {\bibfield  {journal} {\bibinfo  {journal}
		{New Journal of Physics}\ }\textbf {\bibinfo {volume} {16}},\ \bibinfo
	{pages} {112002} (\bibinfo {year} {2014})}\BibitemShut {NoStop}%
\end{thebibliography}
\end{document}